\newif\iffulledition % Boolean to comment stuff not required in reduced edition
\newif\ifnoauthor % Boolean to comment author information

\fulleditiontrue % This means that it is full edition 

\documentclass[conference]{IEEEtran}
\usepackage{amsmath, amssymb, amsthm}
\usepackage{algorithmic}
\usepackage[noend,lined,linesnumbered]{algorithm2e} 
\makeatletter
\renewcommand{\@algocf@capt@plain}{above}% formerly {bottom}
\makeatother
\usepackage{multirow}
\newcommand\user{\mathcal{U}}
\newcommand\server{\mathcal{S}}
\newcommand\adversary{\mathcal{A}}

\usepackage{graphicx}
\usepackage{xcolor}
\usepackage{subfig}
\usepackage{array}

\let\labelindent\relax
\usepackage{enumitem} % For resuming numbering in enumerate. 
\usepackage{color}

\usepackage{hyperref}
\hypersetup{
   colorlinks=true,
   citecolor=teal
}

\begin{document}
\title{BehavioCog: An Observation Resistant Authentication Scheme$^{*}$\thanks{*This is the full version of the paper with the same title which appears in the proceedings of the twenty-first international conference on Financial Cryptography and Data Security 2017 (FC '17).}}

%\author{\IEEEauthorblockN{Michael Shell}
%IEEEauthorblockA{Georgia Institute of Technology\\
%someemail@somedomain.com}
%\and
%\IEEEauthorblockN{Homer Simpson}
%\IEEEauthorblockA{Twentieth Century Fox\\
%homer@thesimpsons.com}
%\and
%\IEEEauthorblockN{James Kirk\\ and Montgomery Scott}
%\IEEEauthorblockA{Starfleet Academy\\
%someemail@somedomain.com}}


\author{\IEEEauthorblockN{Jagmohan Chauhan\IEEEauthorrefmark{1}\IEEEauthorrefmark{2},
Benjamin Zi Hao Zhao\IEEEauthorrefmark{1}\IEEEauthorrefmark{2},
Hassan Jameel Asghar\IEEEauthorrefmark{2},\\ 
Jonathan Chan\IEEEauthorrefmark{2} and
Mohamed Ali Kaafar\IEEEauthorrefmark{2}}
\IEEEauthorblockA{\IEEEauthorrefmark{1}School of Electrical Engineering and Telecommunications\\
UNSW, Sydney, Australia.}
\IEEEauthorblockA{\IEEEauthorrefmark{2}Data61, CSIRO,
Sydney, Australia.\\ Email: {\{jagmohan.chauhan, ben.zhao, hassan.asghar, jonathan.chan, dali.kaafar\}@data61.csiro.au}}}


\IEEEoverridecommandlockouts
%\makeatletter\def\@IEEEpubidpullup{9\baselineskip}\makeatother
%\IEEEpubid{{}}
%\IEEEpubid{\parbox{\columnwidth}{
%Permission to freely reproduce all or part
%    of this paper for noncommercial purposes is granted provided that
%    copies bear this notice and the full citation on the first
%    page. Reproduction for commercial purposes is strictly prohibited
%    without the prior written consent of the Internet Society, the
%    first-named author (for reproduction of an entire paper only), and
%    the author's employer if the paper was prepared within the scope
%    of employment.  \\
  %  NDSS '16, 21-24 February 2016, San Diego, CA, USA\\
   % Copyright 2016 Internet Society, ISBN 1-891562-41-X\\
    %http://dx.doi.org/10.14722/ndss.2016.23xxx
%}
%\hspace{\columnsep}\makebox[\columnwidth]{}}
\graphicspath{ {images/} }

\maketitle

\begin{abstract}
We propose that by integrating behavioural biometric gestures---such as drawing figures on a touch screen---with challenge-response based cognitive authentication schemes, we can benefit from the properties of both. On the one hand, we can improve the usability of existing cognitive schemes by significantly reducing the number of challenge-response rounds by (partially) relying on the hardness of mimicking carefully designed behavioural biometric gestures. On the other hand, the observation resistant property of cognitive schemes provides an extra layer of protection for behavioural biometrics; an attacker is unsure if a failed impersonation is due to a biometric failure or a wrong response to the challenge. We design and develop an instantiation of such a ``hybrid'' scheme, and call it BehavioCog. To provide security close to a 4-digit PIN---one in 10,000 chance to impersonate---we only need two challenge-response rounds, which can be completed in less than 38 seconds on average (as estimated in our user study), with the advantage that unlike PINs or passwords, the scheme is secure under observation.

%The scheme is applicable to any device equipped with a touch screen including smart phones, tablets, smart watches, smart glasses and ATMs.
\end{abstract}

\section{Introduction}
In Eurocrypt 1991~\cite{mat-imai}, Matsumoto and Imai raised an intriguing question: Is it possible to authenticate a user when someone is observing the authentication process? It is easy to see that passwords, PINs or graphical patterns are not secure under this threat model, as the observer can readily impersonate the user after observing one authentication session. Consider for a moment the benefits if an observation resistant authentication scheme existed. We could unlock our smartphones in front of others, authenticate to an ATM with a queue of people behind us, and log on to internet banking in a crowded subway, all without worrying about our authentication credentials being stolen. Unfortunately, more than two decades and a half later, this still remains an open problem. Most proposed solutions to this problem are a form of shared-secret challenge-response type authentication protocols relying on human cognitive abilities, variously called human identification protocols~\cite{Hopper}, cognitive authentication schemes~\cite{CAS} or leakage-resilient password systems~\cite{yan2012limitations}. We will use the term cognitive schemes to refer to them. The idea of using cognitive schemes in part originates from the cryptographic notion of identification protocols, but one where the role of the prover is performed by the human user, and therefore some computations in the protocol have to be done \emph{mentally} by the user. This requirement imposes a design constraint: to make the system usable the cognitive load due to mental computations should be minimized. 

This not only means that only elementary computations should be used, but also that the size of the challenges should be small. But by far the most crippling design constraint is that the response space cannot be too large. If $R$ denotes the response space, and $|R|$ its size, then $|R|$ typically ranges between $2$ and $10$~\cite{Hopper, sechci, CAS, APW}. Observe that anyone can guess the response to a challenge with probability $|R|^{-1}$, even without knowing the secret. Consequently, the number of challenges (or rounds) per authentication session needs to be increased, which in turn increases the authentication time. Let us illustrate this with some examples. To achieve a security equivalent to (guessing) a six digit PIN, i.e., $10^{-6}$, the (low complexity) cognitive scheme (CAS) proposed by Weinshall~\cite{CAS} requires 11 rounds with a total authentication time of 120 seconds, the HB protocol from Hopper and Blum~\cite{Hopper} requires 20 rounds and 660 seconds to authenticate~\cite{yan2012limitations}, and the Foxtail protocol from Li and Shum~\cite{sechci} requires 20 rounds with a cumulative time of 212 seconds~\cite{yan2012limitations}. An authentication time between 10 to 30 seconds per round is perhaps acceptable, since cognitive schemes provide strong security under observation. Thus, if we could significantly reduce the number of rounds, there is a case to use cognitive authentication schemes. 

%In this paper, we propose one way to reduce the number of rounds. 
Our idea is to seek help from gesture-based behavioural biometrics by coupling them with the response space of cognitive schemes. More specifically, we (publicly) map the response space $R$ to a set of $|R|$ different gesture-based \emph{symbols} (such as words or figures) rendered by the user, say, on the touch screen of a smartphone. A classifier decides whether these symbols belong to the target user. We could tune the classifier to achieve a true positive rate (TPR) close to $1$, while giving it some leverage in the false positive rate (FPR), say 0.10. We can now see how we can reduce the number of rounds of the cognitive scheme. Suppose $|R| = 4$ in the cognitive scheme, such as CAS~\cite{CAS}. If we map each of the four responses to four different symbols with an average FPR of 0.10, then the probability of randomly guessing the response to a challenge can be estimated as $\text{FPR} \times |R|^{-1} = 0.025$. Thus, only $4$ rounds instead of 11 will make the guess probability lower than the security of a $6$-digit PIN; an improvement of more than half even with a fairly high FPR. The idea also prevents a possible attack on standalone behavioural biometric based authentication. Minus the cognitive scheme, an imposter can use the behavioural biometric system as an ``oracle'' by continuously adapting its mimicking of the target user's gestures until it succeeds. Integrated with a cognitive scheme, the imposter is unsure whether a failed attempt is due to a biometric error or a cognitive error, or both. The benefit appears mutual. 

Combining the two authentication approaches into a ``hybrid'' scheme is not as simple as it sounds. First, in order to preserve the observation resistance property, the behavioural biometric symbols should be hard to mimic even after an observer has seen users render the symbols on the screen. Behavioural biometric authentication schemes using simple gestures, such as swipes, have been shown to be susceptible to mimicry attacks~\cite{hassan-khan}. Other schemes that employ more complex gestures~\cite{Nguyen, Sherman}, such as free-hand drawings, only tackle shoulder-surfing as opposed to an adversary which has unrestricted access to video transcripts (see Section~\ref{sec:rw}). 
%Furthermore, the said biometric scheme should also be able to produce good accuracy with a small number of training samples, or else the registration phase can beef up to an impractical level. 
%Unable to find a suitable candidate from literature, we propose a new behavioural biometric scheme that is tested against video-based attacks. 
Second, the cognitive scheme CAS~\cite{CAS} used for illustration above is shown to be insecure~\cite{wagner}. Other proposed schemes from literature also lack both the security and usability attributes we desire from a suitable candidate, such as resistance to known attacks, suitability to devices with small displays and low cognitive load in computing responses to challenges (see Section~\ref{sec:rw}). 
%Thus, new constructions for both the cognitive and behavioural biometric schemes are needed.
%This meant that we needed to come up with our own cognitive scheme as well. 

This leads to the other main contributions of our work:

\begin{itemize}
\item We propose a new gesture based behavioural biometric scheme that employs a set of words constructed from certain letters from the English alphabet (e.g., \emph{b},\emph{f},\emph{g},\emph{x},\emph{m}). Previous results from the field of psychology suggest that these letters when written cursively are harder to write in terms of writing pressure and time~\cite{kao1983control, van2011review}. We hypothesized that being harder to write, words constructed from these letters would show more inter-user variation while simultaneously being hard to mimic. Our results indicate that such words are indeed harder to mimic than other words and figures (drawings). 
\iffulledition
The average FPR of these words was 0.0 against random attacks and even after video based observation attacks it went up to only 0.05.
\else
Even under video based observation attacks, the average FPR of these words was 0.05.
\fi
Our proposed behavioural biometric scheme is by itself a contribution, and can be further explored to see its potential as a standalone system secure under observation. 
% Our feature selection methodology shows that features such as writing speed, angle of writing and height are important features and hard to mimic.

%In order for the resulting ``hybrid'' scheme to be secure against observation, the behavioural biometric symbols should be hard to mimic for an observer. We first design symbols based on their likely resistance to observation through a glance at the literature, and then identify those symbols and corresponding features that are hard to mimic through actual mimicking studies. 

\item We propose a new cognitive authentication scheme inspired from the HB protocol~\cite{Hopper} and the Foxtail protocol~\cite{sechci, linearization}. The scheme was designed to balance usability and resistance to known attacks on cognitive authentication schemes. We provide a thorough security analysis of the scheme and suggest parameter sizes against adversarial resources such as time, memory and number of observations. The scheme can be thought of as a contrived version of learning with noisy samples, where the noise is partially a function of the challenge. The generalized form of the resulting scheme is conjectured to resist around $|R| \times n$ number of rounds against computationally efficient attacks, where $n$ is the size of the problem.  

\item We combine the above two into a hybrid authentication scheme which we call BehavioCog and implement it as an app for Android powered smartphones. The app is configurable; parameter sizes of both the cognitive (challenge size, secret size, response space, etc.) and behavioural biometric (symbols, amount of training, etc.) components can be tuned at set up. 
%The registration phase is also tunable giving the option to increase or decrease the amount of training given to with increasing time and complexity. 
The app partially guided setting parameter sizes for the BehavioCog scheme, such as the size of challenges and length/size of symbols, based on what was viewable and writeable on the smartphone touch screen. Our cognitive scheme is implemented using a set of emojis (a set of ``smileys'' used for digital communication). However, emojis can be replaced by any other set of images or characters.

\item We carry out a user study with 41 different users and do an extensive analysis of the usability, security and repeatability of our scheme. We video recorded the authentication sessions to simulate a hidden camera and allowed users to act as attackers by impersonating a target user providing them with video transcripts and unrestricted control over the playback. None of the attacks were successful in our scheme (with two rounds in one authentication session). Our results also give separate statistics, such as authentication time and errors, for the cognitive and biometric components. Results indicate that the time taken to compute the cognitive challenge can be as low as 12 seconds (on average) and 6 seconds (on average) to submit the response through a biometric symbol. 
%Our results show that we can achieve an overall success rate of 85\%. All errors were cognitive errors, meaning that the biometric errors were none. A week later the cognitive errors increased, while biometric errors remained low (8\%). We extended the training to include a trick from cognitive psychology (showing people see images for a few seconds followed by a brief cool-off period with a blank screen). We achieved a success rate of 92\% after that. 
\end{itemize}

In terms of overall results, we find that the average authentication time for each round could go as low as 19 seconds and  we can achieve security comparable to a 4-digit PIN within just 2 rounds, and close to 6-digit PIN with 3 rounds, even under observation attacks. The error rate was relatively high for the cognitive scheme with the best case being 15\%.\footnote{In fact, we show that by carefully designing the training module we can reduce the error to 14\% even after a gap of a week. See Section~\ref{sub:results:phase3}} However, the study was intentionally carried out with a high number of pass-emojis (cognitive secret), 14 to be precise, since smaller sized graphical passwords (say 3 to 5) have already been demonstrated to be recognizable \cite{Dhamija, Hayashi}. Moreover, our scheme also allows smaller sizes of pass-emojis with a trade off that a lower number of observations can be resisted.

We do not claim that our idea completely solves the problem raised by Matsumoto and Imai, but believe it to be a step forward towards that goal.\footnote{Also see our discussion in Section~\ref{sec:discuss}.} In addition, we believe that our proposal can potentially revive interest in research on cognitive authentication schemes and their application as a separate factor in multi-factor authentication schemes. 
We have organized the rest of the paper as follows. Section~\ref{sec:overview} contains overview of the BehavioCog scheme and the threat model followed by the description of the cognitive scheme in Section~\ref{sec:cognitive} and the behavioural biometric scheme in Section~\ref{sec:biometrics}. Implementation details are given in Section~\ref{sec:implement}. Sections~\ref{sec:experiments} and \ref{sec:results} detail the user study and its results, respectively. Related work is summarized in Section~\ref{sec:rw}. We discuss some limitations and identify areas of future research in Section~\ref{sec:discuss} and conclude the paper with Section~\ref{sec:conclude}.

\section{Overview of BehavioCog}
\label{sec:overview}
\iffulledition
We begin with defining authentication schemes and the adversarial model, followed by the overview of our BehavioCog scheme.
\fi
\subsection{Preliminaries}
\subsubsection{Authentication Schemes}
An \emph{authentication} scheme consists of two protocols: \emph{registration} and \emph{authentication}, between two parties: the prover and the 
\iffulledition
verifier~\cite{schoenmakers}. 
\else
verifier.
\fi
In this paper, we assume the prover to be a user $\mathcal{U}$, and the verifier to be an authentication service $\mathcal{S}$. A \emph{shared-secret challenge-response} authentication scheme is an authentication scheme in which $\user$ and $\server$ share a secret $x$ from a secret space $X$ during registration, and the authentication phase is as follows:
\iffulledition
\begin{algorithm}[!h]
\SetAlgoLined
\SetCommentSty{mycommfont}
\SetAlCapSkip{1em}
\DontPrintSemicolon{}
\SetKwInOut{Input}{Input}
\let\oldnl\nl
\newcommand{\nonl}{\renewcommand{\nl}{\let\nl\oldnl}}
\For{$\gamma$ rounds}{
$\server$ sends a challenge $c \in C$ to $\user$. \;
$\user$ sends the response $r = f(x, c)$ to $\server$.\; 
}
\textbf{if} all $\gamma$ responses are correct $\server$ accepts $\user$, \textbf{else} it rejects $\user$. 
\end{algorithm}
\else
For $\gamma$ rounds, $\server$ sends a challenge $c \in C$ to $\user$, who sends the response $r = f(x, c)$ back to $\server$. If all $\gamma$ responses are correct $\server$ accepts $\user$, else it rejects $\user$.
\fi
Here, $C$ is the challenge space, and $r$ belongs to a response space $R$. The function $f: X \times C \rightarrow R$ is what we shall refer to as a cognitive function in this paper. Note that apart from the selected secret $x \in X$, everything else is public. Since $x$ is shared, $\server$ can compute $f$ itself to check if the responses are correct. A challenge and a response from the same round shall be referred to as a challenge-response pair. An authentication session, therefore, consists of $\gamma$ challenge-response pairs as above. In practice, we assume $\user$ and $\server$ interact via the user's device (e.g., a smartphone) and the function $f$ needs to be computed mentally by $\user$. 

\subsubsection{Adversarial Model}
We assume a passive adversary $\adversary$ who can observe one or more authentication sessions between $\user$ and $\server$. In other words, $\adversary$ can see challenge-response pairs exchanged between the two parties. The goal of $\adversary$ is to impersonate $\user$ by initiating a new session with $\server$ and making it accept $\adversary$ as $\user$. $\adversary$ can succeed in doing so either by learning the secret $x$ or by correctly guessing $\gamma$ responses. 
\iffulledition
The probability of the latter approach is $|R|^{-\gamma}$. Thus, $\gamma$ can be set such that this quantity is below the desired security level. 
\fi
In practice, we assume that $\adversary$ can observe the screen of the device used by $\user$ (on which the challenges are displayed and the responses are entered). This can be done either via shoulder-surfing (simply by looking over $\user$'s shoulder) or via a video recording using a spy camera. 
\iffulledition
Clearly, the latter attack is more powerful, and a system secure under this attack is arguably also secure against direct shoulder-surfing. 
\fi
We assume that $\user$'s device is secure.\footnote{The original threat model from Matsumoto and Imai also assumes that the device as well as the communication channel between the device and $\server$ are insecure. Our threat model is slightly restricted.}
\subsection{The BehavioCog Scheme}
The main idea of our hybrid authentication scheme is as follows. Instead of sending the response $r$ to a challenge $c$ from $\server$, $\user$ renders a \emph{symbol} corresponding to $r$, and this rendered symbol is then sent to $\server$. In practice, this is implemented as $\user$ rendering the symbol on the touch screen of the user's device. More specifically, we assume a set of symbols denoted $\Omega$, e.g., a set of words in English,\footnote{For a word, rendering means writing.} where the number of symbols equals the number of responses $|R|$. Each response $r \in R$ is mapped to a symbol in $\Omega$. The symbol corresponding to $r$ shall be represented by $\text{sym}(r)$. Upon receiving the rendering of $\text{sym}(r)$, $\server$ first checks if the rendered symbol ``matches'' a previously stored rendering from $\user$ and then checks if the response $r$ is correct by computing $f$. If the answer to both is yes in each challenge-response round, $\server$ accepts the user. Two observations are in order. First, the registration phase now needs to collect sample renderings for each symbol in $\Omega$, which we refer to as \emph{templates}. Second, in order to compare a rendering against the stored templates we need a classifier $D$ that can decide if it matches any of the $|\Omega|$ templates or not. 

The scheme is described via a series of protocols described in Figure~\ref{fig:scheme}. To explain the protocols, we begin by detailing the cognitive scheme first. Assume a global pool of $n$ objects (e.g., images or emojis). A secret $x \in X$ is a $k$-element subset of the global pool of objects. Thus, $|X| = \binom{n}{k}$. Each object of $x$ is called a pass-object, and the remaining $n - k$ objects are called decoys. The challenge space $C$ consists of pairs $c = (a, w)$, where $a$ is an $l$-element sequence of objects from the global pool, and $w$ is an $l$-element sequence of integers from $\mathbb{Z}_d$, where $d \ge 2$. Members of $w$ shall be called weights. The $i$th weight in $w$ is denoted $w_i$ and corresponds to the $i$th element of $a$, i.e., $a_i$. The notation $c \in_U C$, which means sampling a random $l$-element sequence of objects $a$ and a random $l$-element sequence of weights $w$. The cognitive function $f$ is defined as
\begin{equation}
\label{eq:cog}
f(x, c) =
\begin{cases}
    \sum_{i \mid a_i \in x} w_i \bmod d, & \text{if $x \cap a \ne \emptyset$} \\
    r \in_U \mathbb{Z}_d, & \text{if $x \cap a = \emptyset$}.
  \end{cases}
\end{equation}
That is, sum all the weights of the pass-objects in $c$ and return the answer modulo $d$. If no pass-object is present then return a random element from $\mathbb{Z}_d$. The notation $\in_U$ means sampling uniformly at random. It follows that the response space $R = \mathbb{Z}_d$ and $|R| = d$. Now, let $\Omega$ be a set of $d$
\iffulledition
symbols. Later on, we shall show different instantiations of $\Omega$. For now, we can consider $\Omega$ to be the set of English words of integers in $\mathbb{Z}_d$, i.e., $\mathtt{zero}$, $\mathtt{one}$, $\mathtt{two}$, and so on.
\else
symbols, e.g., the words $\mathtt{zero}$, $\mathtt{one}$, $\mathtt{two}$, and so on.
\fi
The mapping $\text{sym}: \mathbb{Z}_d \rightarrow \Omega$ is the straightforward lexicographic mapping. Note that we do not require this mapping to be a secret. We assume a $(d + 1)$-classifier $D$ which when given as input the templates of all symbols in $\Omega$, and a rendering of some symbol purported to be from $\Omega$, outputs the corresponding symbol in $\Omega$ if the rendering matches any of the symbol templates. If no match is found, $D$ outputs ``none.'' $D$ needs a certain number of renderings of each symbol to build its templates. We denote this number by $t$. For instance, $t = 3, 5$ or $10$. Note that since $\server$ rejects whenever a rendering does not match the template, we need to tune $D$ to have a near 100 percent TPR to avoid errors. The specific classifier we use is based on the dynamic time warping (DTW) 
\iffulledition
algorithm~\cite{dtw-history, dtw-muller}. 
\else
algorithm~\cite{dtw-muller}. 
\fi
We postpone the details of the classifier till Section~\ref{sec:biometrics}. Figure~\ref{fig:scheme} describes the above process concisely via protocols for set up, registration and authentication. 

\makeatletter
\newcommand{\removelatexerror}{\let\@latex@error\@gobble}
\makeatother
\begin{figure*}[!ht]
\centering
{\footnotesize
\begin{minipage}[t]{0.30\linewidth}
\removelatexerror% Nullify \@latex@error
\begin{algorithm}[H]
\SetAlgorithmName{1}{}
\SetAlgoLined
\SetAlCapSkip{1em}
\DontPrintSemicolon{}
\let\oldnl\nl% Store \nl in \oldnl
\newcommand{\nonl}{\renewcommand{\nl}{\let\nl\oldnl}}% Remove line number for one line
\nonl\TitleOfAlgo{\textbf{Setup.}}
$\mathcal{S}$ publishes parameters $n$, $k$, $l$ and $d$ (e.g., $n = 180$, $k = 14$, $l = 30$, $d = 5$). \;
$\mathcal{S}$ publishes the global pool of $n$ objects (e.g., emojis). \;
$\mathcal{S}$ publishes a set of $d$ symbols $\Omega$ (e.g., words).\;
$\mathcal{S}$ publishes the map $\text{sym}$ from $\mathbb{Z}_d$ to $\Omega$. \;
$\mathcal{S}$ publishes the (untrained) classifier $D$. \;
$\mathcal{S}$ publishes $t$, the required number of renderings per symbol to create templates (e.g., $t = 10$).  \;
\end{algorithm}
%}
\end{minipage}%
\begin{minipage}[t]{0.25\linewidth}
\removelatexerror% Nullify \@latex@error
\begin{algorithm}[H]
\SetAlgorithmName{2}{} 
\SetAlgoLined
\SetAlCapSkip{1em}
\DontPrintSemicolon{}
\let\oldnl\nl% Store \nl in \oldnl
\newcommand{\nonl}{\renewcommand{\nl}{\let\nl\oldnl}}% Remove line number for one line
\nonl\TitleOfAlgo{\textbf{Registration.}}
$\user$ and $\server$ share a secret $x \in X$. \;
For each symbol in $\Omega$, $\user$ sends $t$ renderings to $\server$.\;
For each symbol in $\Omega$, $\server$ trains $D$ on the $t$ renderings to obtain $\user$'s template.\;
The secret consists of $x$ and the $d$ templates.\;
\end{algorithm}
%}
\end{minipage}%
\begin{minipage}[t]{0.45\linewidth}
%\subfloat[$k = 5, n = 20$ \label{fig:k5n21}]{\scalebox{0.29}
\removelatexerror% Nullify \@latex@error
\begin{algorithm}[H]
\SetAlgorithmName{3}{} 
\SetAlgoLined
\SetAlCapSkip{1em}
\DontPrintSemicolon{}
\let\oldnl\nl% Store \nl in \oldnl
\newcommand{\nonl}{\renewcommand{\nl}{\let\nl\oldnl}}% Remove line number for one line
\nonl\TitleOfAlgo{\textbf{Authentication.}}
$\server$ sets $\text{err} = 0$.\; 
\For{$\gamma$ rounds}{
	$\server$ samples $c = (a, w) \in_U C$ and sends it to $\user$.\;
	$\user$ computes $r = f(x, c)$.\; 
	$\user$ renders the symbol $\text{sym}(r)$, and sends it to $\server$.\; 
	$\server$ runs $D$ on the rendering.\;
	\If{$D$ outputs ``none''}{ 
		$\server$ sets $\text{err} = 1$.\;
	}
	\Else{
		$\server$ obtains $r$ corresponding to the symbol output by $D$.\;
		\If{$x \cap a \ne \emptyset$ and $r \ne f(x, c)$}{
			$\server$ sets $\text{err} = 1$.\;
		}
	}
}
\textbf{If} $\text{err} = 1$, $\server$ rejects $\user$; otherwise it accepts $\user$.\;
\end{algorithm}
%}
\end{minipage}%
}
\caption{The setup, registration and authentication protocols of BehavioCog.\label{fig:scheme}}
\end{figure*}

Since the set up and registration protocols are straightforward, we only briefly describe the authentication protocol here. $\server$ initializes an \emph{error} flag to 0 (Step 1). Then, for each of the $\gamma$ rounds, $\server$ sends $c = (a, w) \in_U C$ to $\user$ (Step 3). $\user$ computes $f$ according to Eq.~\ref{eq:cog}, and obtains the response $r$ (Step 4). $\user$ gets the symbol to be rendered through $\text{sym}(r)$, and sends a rendering of the symbol to $\server$ (Step 5). Now, $\server$ runs the trained classifier $D$ on the rendered symbol (Step 6). If the classifier outputs ``none,'' $\server$ sets the error flag to 1 (Step 8). Otherwise, $D$ outputs the symbol corresponding to the rendering. Through the inverse map, $\server$ gets the response $r$ corresponding to the symbol (Step 10). Now, if $x \cap a = \emptyset$, i.e., none of the pass-objects are in the challenge, then any response $r \in \mathbb{Z}_d$ is valid, and therefore $\server$ moves to the next round. Otherwise, if $x \cap a \ne \emptyset$, $\server$ further checks if $r$ is indeed the correct response by computing $f$ (Step 11). If it is incorrect, $\server$ sets the error flag to 1 (Step 12). Otherwise, if the response is correct, $\server$ moves to the next round. If after the end of $\gamma$ rounds, the error flag is 0, then $\server$ accepts $\user$, otherwise it rejects $\user$ (Step 13).  

\section{The Cognitive Scheme}
\label{sec:cognitive}
Our proposed cognitive scheme can be thought of as an amalgamation of the HB scheme based on the learning parity with noise (LPN) problem~\cite{Hopper}, and the Foxtail scheme (with window)~\cite{sechci, linearization}. More specifically, we use the idea of using an $l$-element challenge from the Foxtail with window scheme. However instead of using the Foxtail function, which maps the sum of integers modulo $d = 4$, to $0$ if the sum is in $\{0, 1\}$, and $1$ otherwise, we output the sum itself as the answer. The reason for that is to reduce the number of rounds, i.e., $\gamma$, for a required security level (the success probability of random guess is $\frac{1}{2}$ in one round of the Foxtail scheme). Now if we allow the user to output $0$ in case none of its pass-objects are present in a challenge, the probability distribution of the output of $f$ is skewed towards $0$, which makes the scheme susceptible to a statistical attack proposed by Yan et al.~\cite{yan2012limitations} outlined in Section~\ref{sub:fa}. In order to circumvent this, we ask the user to output a random response from $\mathbb{Z}_d$ whenever none of the pass-objects are present. This fix makes the scheme immune to the statistical attack. We shall provide more details in Section~\ref{sub:fa}. Due to the random response, we can say that the resulting scheme adds noise to the samples (challenge-response pairs) collected by $\adversary$, somewhat similar in spirit to HB. The difference is that in our case, the noise is (partially) a function of the challenge, whereas in HB the noise is independently generated with a fixed probability and added to the sum. Having laid out the main idea behind the cognitive scheme, we now discuss its security in detail.

\subsection{Random Guess Attack}
\label{sub:rg}
\iffulledition
In principle, two kinds of random guess attacks are possible: randomly guessing the secret and then computing the response according to $f$, or randomly guessing the response. The success probability of the first form is proportional to $\binom{n}{k}^{-1}$, which is negligible if $n$ and $k$ are moderately large. We therefore only consider the second form of random guess. 
\fi
Let $p_{\text{RG}}$ denote the success probability of a random guess. This probability is conditioned on the event $a \cap x$ being $empty$ or not. Since this event shall be frequently referred to in the text, we give it a special name: the \emph{empty case}. Now the probability that $i$ pass-objects are present in $a$, is given by 
\iffulledition
the probability mass function of the hypergeometric distribution 
\[
\mathbb{P}\left[| a \cap x | = i \right] = \frac{\binom{k}{i}\binom{n-k}{l-i}}{\binom{n}{l}},
\]
\else
$\mathbb{P}\left[| a \cap x | = i \right] = {\binom{k}{i}\binom{n-k}{l-i}}/{\binom{n}{l}}$,
\fi
from which it follows that
\iffulledition
\[
\mathbb{P}\left[| a \cap x | = 0 \right] \doteq p_0 = \frac{\binom{n-k}{l}}{\binom{n}{l}}.
\]
We shall use the notation $\doteq$ when defining a variable. Thus, 
\[
p_{\text{RG}} = p_0 + (1 - p_0) \frac{1}{d}.
\]
\else
$\mathbb{P}\left[| a \cap x | = 0 \right] \doteq p_0 = {\binom{n-k}{l}}/{\binom{n}{l}}$. 
We shall use the notation $\doteq$ when defining a variable. Thus, 
$
p_{\text{RG}} = p_0 + (1 - p_0) \frac{1}{d}.
$
\fi
%For $(d, k, l, n) = (5, 10, 30, 180)$, we have $p_{\text{RG}} \approx 0.323$ and for $(d, k, l, n) = (5, 14, 30, 180)$, we have $p_{\text{RG}} \approx 0.256$

\subsection{Brute Force Attack and Information Theoretic Bound}
This attack is only possible after $\adversary$ has observed $m > 0$ challenge-response pairs (or samples) corresponding to successful authentication sessions. Before observing any samples, i.e., $m = 0$, all possible $\binom{n}{k}$ subsets are possible candidates of the target secret $x$. We denote a candidate by $y$, where quite possibly $y = x$. After observing one sample, the probability that a $y \in X$ is still a candidate for the secret $x$ is given by
\iffulledition
\[
p_0 + (1 - p_0) \frac{1}{d},
\] 
which is the same as the probability of a successful random guess. 
\else
$p_0 + (1 - p_0) \frac{1}{d}$.
\fi
This follows because if $a \cap y = \emptyset$, i.e., if it is the empty case, then $y$ is trivially consistent with any response. On the other hand, if $a \cap y \ne \emptyset$, then the probability that the response from $y$ matches that of $x$ is $\frac{1}{d}$. Thus, we expect 
\iffulledition
\[
\left( p_0 + (1 - p_0) \frac{1}{d} \right)^m \binom{n}{k},
\]
\else
$\left( p_0 + (1 - p_0) \frac{1}{d} \right)^m \binom{n}{k}$
\fi
subsets in $X$ to still remain as candidates for $x$ after observing $m$ challenge-response pairs. Equating the above to $1$, gives us
\iffulledition
\[
m \doteq m_{\text{it}} = - \frac{\log_2{\binom{n}{k}}}{\log_2(p_0 + (1 - p_0) \frac{1}{d})}.
\]
\else
$m \doteq m_{\text{it}} = - {\log_2{\binom{n}{k}}}/{\log_2(p_0 + (1 - p_0) \frac{1}{d})}$.
\fi
We call $m_{\text{it}}$, the information theoretic bound on $m$. This is the least (expected) number of samples needed to be observed to obtain a unique candidate for the secret. 
%The complexity of the brute force attack mentioned above is $\binom{n}{k}$, which is $\approx 2^{53}$ when $(k, n) = (10, 180)$, $\approx 2^{68}$ when $(k, n) = (14, 180)$ and $\approx 2^{81}$ when $(k, n) = (18, 180)$. $m_{\text{it}}$ for  $(d, k, l, n) = (5, 14, 30, 180)$ is $\approx 34$. 

\subsection{Meet-in-the-Middle Attack}
This attack~\cite{Hopper} works by first computing $\frac{k}{2}$-sized subsets of $X$ on each of the $m$ observed challenge-response pairs, and storing the $m$-element response string together with the subset in a hash table. After that, for each possible ``intermediate'' response string in $\mathbb{Z}^m_d$, and for each $\frac{k}{2}$-sized subsets of $X$ we compute the final response string of $m$-elements. If this response string matches at least $m(1 - p_0)$ responses\footnote{i.e., the expected number of samples that do not belong to the empty case.} in the response string of the target secret $x$, we insert the intermediate response string together with the corresponding $\frac{k}{2}$-sized subset in the same hash table. Any collision in the hash table marks a possible candidate for $x$ (by combining the two $\frac{k}{2}$-sized subsets). In a post-processing step, we can check whether a candidate $y$ is consistent by checking if the $mp_0$ fraction of incorrect responses correspond to the empty case. The time and space complexity of this attack is $\binom{n}{k/2}$. 
\iffulledition
We note that there is another meet-in-the-middle attack shown in~\cite{APW} based on Coppersmith's baby-step giant-step algorithm~\cite{stinson}, of time complexity $\binom{n/2}{k/2}$, but with the same space complexity of $\binom{n}{k/2}$.
\fi

\subsection{Frequency Analysis}
\label{sub:fa}
Frequency analysis is an attack proposed by Yan et al.~\cite{yan2012limitations}\footnote{We borrow the term frequency analysis from~\cite{count-counts}.} which could be done either independently or dependent on the response. In response-independent frequency analysis (RIFA), a frequency table of $\delta$-tuples of objects is created, where $1 \le \delta \le k$. If a $\delta$-tuple is present in a challenge, its frequency is incremented by $1$. After gathering enough challenge-response pairs, the tuples with the highest or lowest frequencies may contain the $k$ secret objects if the challenges are constructed with a skewed distribution. In the response-dependent frequency analysis (RDFA), the frequency table contains frequencies for each possible response in $\mathbb{Z}_d$, and the frequency of a $\delta$-tuple is incremented by $1$ in the column corresponding to the response (if present in the challenge). 

First, note that our cognitive scheme is resistant to RIFA since the challenges are drawn uniformly at random without considering pass or decoy objects. This follows from Lemma 17 in \cite{count-counts}. To see that RDFA is also not applicable, define the indicator random variable
\iffulledition
\begin{equation*}
I(x') =
\begin{cases}
    1, & \text{if each element of $x'$ is in $a$} \\
    0, & \text{otherwise},
  \end{cases}
\end{equation*}
\else
$I(x')$ which is $1$ if $x' \in a$, 
\fi
where $x' \subseteq x \in X$. We define a similar indicator random variable $I(y')$ for $y' \subseteq y \in X^{n - k}$, where $X^{n - k}$ denotes the set of $n - k$ decoy objects. Now for RDFA to be inapplicable we should have
\iffulledition
\[
\mathbb{P}\left[ I(x') = b \mid r = i \right] = \mathbb{P}\left[ I(y') = b \mid r = i \right], 
\] 
\else 
$\mathbb{P}\left[ I(x') = b \mid r = i \right] = \mathbb{P}\left[ I(y') = b \mid r = i \right]$, 
\fi
for $i \in \mathbb{Z}_d$, $b \in \{0, 1\}$ and $|x'| = |y'|$. Using Baye's rule
\[
\mathbb{P}\left[ I(x') = b \mid r = i \right] = \frac{\mathbb{P}\left[ r = i \mid I(x') = b   \right] \mathbb{P}\left[ I(x') = b \right]}{\mathbb{P}\left[ r = i \right]}.
\]
Now, 
\iffulledition
\[
\mathbb{P}\left[ r = i \right] = p_0\cdot \frac{1}{d} + (1 - p_0) \frac{1}{d} = \frac{1}{d}.
\]
Also, from Lemma 17 in \cite{count-counts}
\[
\mathbb{P}\left[ I(x') = 1 \right] = \mathbb{P}\left[ I(y') = 1 \right] = \frac{\binom{n - \delta}{l - \delta}}{\binom{n}{l}},
\]
where $\delta \doteq |x'| = |y'|$. From the above, it follows that $\mathbb{P}\left[ I(x') = 0 \right] = \mathbb{P}\left[ I(y') = 0 \right] $. Now, 
\[
\mathbb{P}\left[ r = i \mid I(y') = b   \right] = \mathbb{P}\left[ r = i \right]  = \frac{1}{d},
\]
since the responses are not dependent on the decoy objects. Finally, we see that 
\[
\mathbb{P}\left[ r = i \mid I(x') = 1   \right] = \frac{1}{d},
\]
\else
$
\mathbb{P}\left[ r = i \right] = p_0\cdot \frac{1}{d} + (1 - p_0) \frac{1}{d} = \frac{1}{d}.
$
Also, from Lemma 17 in \cite{count-counts}
$
\mathbb{P}\left[ I(x') = 1 \right] = \mathbb{P}\left[ I(y') = 1 \right] = {\binom{n - \delta}{l - \delta}}/{\binom{n}{l}},
$
where $\delta \doteq |x'| = |y'|$. From this, it follows that $\mathbb{P}\left[ I(x') = 0 \right] = \mathbb{P}\left[ I(y') = 0 \right] $. Now, 
$
\mathbb{P}\left[ r = i \mid I(y') = b   \right] = \mathbb{P}\left[ r = i \right]  = \frac{1}{d},
$
since the responses are not dependent on the decoy objects. Finally, we see that 
$
\mathbb{P}\left[ r = i \mid I(x') = 1   \right] = \frac{1}{d},
$
\fi
since at least $\delta$ pass-objects are present in the challenge, and the response is the sum modulo $d$, which due to the randomness of weights is distributed uniformly in $\mathbb{Z}_d$. If $I(x') = 0$, there are two possibilities. Either $\delta-1$ or less number of pass-objects are present in the challenge, in which case the response is again uniform in $\mathbb{Z}_d$, or none of the pass-objects are present (empty case). But in the latter case, we ask the user to output a random response in $\mathbb{Z}_d$. Therefore, the probability of observing a response $r = i$  is $\frac{1}{d}$. From this it follows that our scheme is secure against RDFA. 

\subsection{Coskun and Herley Attack}
Since only $l$ objects are present in each challenge, the number of pass-objects present is also less than $k$ with high probability. Let $u$ denote the average number of bits of $x$ used in responding to a challenge. The Coskun and Herley (CH) attack~\cite{ch-attack} states that if $u$ is small, then candidates $y \in X, y \ne x$, that are close to $x$ in terms of some distance metric, will output similar responses to $x$. If we sample a large enough subset from $X$, then with high probability there is a candidate for $x$ that is a distance $\xi$ from $x$. We can remove all those candidates whose responses are far away from the observed responses, and then iteratively move closer to $x$. The running time of the CH attack is at least 
${|X|}/{\binom{\log_2 |X|}{\xi}}$~\cite{ch-attack} where $|X| = \binom{n}{k}$, with the trade off that $m \approx \frac{1}{\epsilon^2}$ samples are needed for the attack to output $x$ with high probability~\cite{ch-bounds, lin-crypt}. The parameter $\epsilon$ is the difference in probabilities that distance $\xi + 1$ and $\xi - 1$ candidates have the same response as $x$. 
\iffulledition
More specifically $\epsilon$ is given by~\cite{ch-bounds}
\[
\epsilon \doteq \left( \frac{\binom{\log_2 |X| - \xi + 1}{\upsilon}}{\binom{\log_2 |X|}{\upsilon}} - \frac{\binom{\log_2 |X| - \xi - 1}{\upsilon}}{\binom{\log_2 |X|}{\upsilon}} \right) \left( 1 - \frac{1}{d} \right),
\]
where $\upsilon$ is the average number of bits of the secret used per challenge, and can be estimated as $\frac{l}{n} \log_2 |X|$~\cite{ch-bounds}.
\fi
As we choose higher values of $\xi$, the complexity of the attack decreases but the probability differences become less prominent, which in turn means that more samples $m$ need to be observed. The optimal value of $\xi$ is when the time complexity is below a threshold, giving us a value of $\epsilon$ from which the number of required samples $m$ can be obtained~\cite{ch-bounds}. 

\subsection{Linearization}
We begin by assigning an order to the $n$ objects in the global pool. We can then represent the secret $x$ as an $n$-element binary vector $\mathbf{x}$ of Hamming weight $k$ (where $x_i = 1$ indicates that object $i$ is present in the secret). Similarly, a challenge $c = (a, w)$ can be represented by the $n$-element binary vector $\mathbf{a}$ of Hamming weight $l$ (indicating the presence of the corresponding object) and the $n$-element vector $\mathbf{w}$ of Hamming weight $\le l$, where $w_i = 0$ if $a_i = 0$. Let $\eta \in_U \mathbb{Z}_d$. Then our cognitive function $f$ can be rewritten as 
\iffulledition
\begin{equation}
\label{eq:linear}
f(\mathbf{x}, \mathbf{c}) = b \mathbf{w} \cdot {\mathbf{x}} + \eta (1 - b) \bmod d,
\end{equation}
\else
$f(\mathbf{x}, \mathbf{c}) = b \mathbf{w} \cdot {\mathbf{x}} + \eta (1 - b) \bmod d$,
\fi
where $b = \text{sgn}(\mathbf{a} \cdot \mathbf{x})$ is the sign function. Now, consider the case $r \doteq f(\mathbf{x}, \mathbf{c}) = 0$. This is possible if $b = 1$ and $\mathbf{w} \cdot {\mathbf{x}} \equiv 0 \bmod d$, or when $b = 0$ and $\eta = 0$. In the latter case, note that $\mathbf{w} \cdot {\mathbf{x}} = 0$ (even without the modulus), and hence trivially $\mathbf{w} \cdot {\mathbf{x}} \equiv  0 \bmod d$. On the other hand, if $r \ne 0$, we again have the possibility that if $b = 1$, $\mathbf{w} \cdot {\mathbf{x}} \equiv r \bmod d$ or if $b = 0$, then $\eta = r$. However, we cannot write the latter as an equation in $\mathbf{x}$ and $\mathbf{w}$ without including the non-zero noise term $\eta$. 

Thus one attack strategy is to keep samples corresponding to a $0$ response to build a system of linear congruences. After $n$ such congruences have been obtained, $\adversary$ can use Gaussian elimination to obtain a unique solution for $\mathbf{x}$, thus obtaining the secret. That is, create the matrix $W$ whose $i$th row corresponds to the weight vector from the $i$th challenge $\mathbf{c}_i$ such that the corresponding response is $0$. This gives us the system of linear congruences $W\mathbf{x} \equiv \mathbf{0} \bmod d$, where $W$ is an $n \times n$ square matrix. Of course, $W$ needs to be a full rank matrix. This can be done by observing a little over $n$ samples (with 0 response), because with high probability a randomly generated $W$ is of full rank if $l$ is large enough~\cite{ch-bounds, omega}. For instance, with $(k, l, n) = (14, 30, 140)$ we found that a fraction 0.29 of the matrices generated had full rank by running a Monte Carlo simulation with 10,000 repetitions. Note that since the response is uniformly distributed in $\mathbb{Z}_d$, we expect to construct $W$ after observing $dn$ challenge-response pairs. Thus, we are discarding all challenges that correspond to a non-zero response. 

Another way of linearization that does not discard any challenges, but requires the observations of the same number of challenge-response pairs, is to introduce $(d-1)n$ new binary variables. We illustrate this using $d = 2$ as an example. Let $\mathbf{w}_i$ denote the $i$th $n$-element weight vector. Then we can form the system
\[ 
 \begin{pmatrix}
  \mathbf{w}_1  & 1 & 0 & \cdots & 0 \\
  \mathbf{w}_2 & 0 & 1 & \cdots & 0 \\
   \vdots  & \vdots & \vdots  & \ddots &\vdots \\
 \mathbf{w}_n & 0 & 0 & \cdots & 1 \\
  \mathbf{w}_{n+1} & 0 & 0 & \cdots & 0 \\
  \vdots  & \vdots & \vdots  & \ddots &\vdots \\
  \mathbf{w}_{2n} & 0 & 0 & \cdots & 0
 \end{pmatrix}  
 \begin{pmatrix}
 x_1 \\
 \vdots \\
 x_n \\
 x_{n+1} \\
 \vdots \\
 x_{2n}
  \end{pmatrix}  
 \equiv 
 \begin{pmatrix}
 1 \\
 \vdots \\
 1 \\
 0 \\
 \vdots \\
 0
  \end{pmatrix} \bmod 2,
\]
where $x_{n+1}, \ldots, x_{2n}$ are $n$ new variables. The above system of equations is obtained by observing $2n$ challenge-response pairs and re-arranging the $0$ and $1$ responses (the top $n$ rows correspond to $r = 1$). Let us call the $2n \times 2n$ matrix, $W$. By construction of the last $n$ columns of $W$, the $2n$ rows of $W$ are linearly independent regardless of the vectors $\mathbf{w}_1, \ldots, \mathbf{w}_n$ as long as the vectors $\mathbf{w}_{n+1}, \ldots, \mathbf{w}_{n}$ remain linearly independent. But we have seen above that this is true with high probability. Hence, we can use Gaussian elimination again to uniquely obtain the secret. To see that the above system is consistent with observation, consider the first row. If it corresponds to the empty case, then by setting $x_{n+1} = 1$ we get the response $1$. On the other hand, if it is not the zero case then $x_{n+1} = 0$ satisfies the equation. Any of the two values of $x_{n+1}$ satisfy the $0$-response rows. Since the responses are generated randomly, we expect to obtain the above system by observing $dn$ challenge-response 
\iffulledition
pairs. Note that if $\user$ were to respond with $0$ in the empty case, then we could obtain a linear system of equations after $n$ challenge-response pairs. The introduction of noise expands the number of required challenge-response pairs to $dn$, an increase by a factor of $d$. Gaussian elimination is by far the most efficient attack on our scheme, and therefore this constitutes a significant gain. 
\subsection{Generalization}
With the exception of Gaussian elimination, all other attacks mentioned above have complexity exponential in one or more variables in $(k, l, n)$. Since the above linearization works after observing $dn$ challenge-response pairs, we believe the problem of finding a polynomial time algorithm in $(k, l, n)$ which uses $m < dn$ number of samples (say $(d-1)n$ samples) from the function described in Eq.~\ref{eq:cog} is an interesting open question.
\else
pairs. Note that if $\user$ were to respond with $0$ in the empty case, then we could obtain a linear system of equations after only $n$ challenge-response pairs. We believe the problem of finding a polynomial time algorithm in $(k, l, n)$ which uses $m < dn$ number of samples (say $(d-1)n$ samples) from the function described in Eq.~\ref{eq:cog} is an interesting open question.
\fi

\subsection{Example Parameter Sizes}
\label{sub:paramcog}
Table~\ref{table:paramcog} shows an example list of parameter sizes for the cognitive scheme. These sizes are obtained by fixing $d = 5$ and changing $k$, $l$ and $n$ such that the random guess probability $p_{\text{RG}}$ is approximately $0.25$. We suggest $d = 5$ as a balance between reducing the number of rounds required, i.e., $\gamma$, and ease of computing $f$. The column labelled $m_{\text{it}}$ is the information theoretic bound on minimum (expected) number of samples needed to uniquely obtain the secret. Thus, the first two suggestions are only secure with $\le m_{\text{it}}$ observed samples. The complexity shown for both the meet-in-the-middle attack (MitM) and Coskun and Herley (CH) attack represents time as well as space complexity. The last column is Gaussian elimination (GE). The required number of samples for GE is calculated as $dn$. For other attacks, we show the minimum number of required samples $m$, such that $m \ge m_{\text{it}}$ and the complexity is as reported. We can think of the last two suggested sizes as secure against an adversary with time/memory resources $\approx 2^{70}/ 2^{40}$ (medium strength) and $\approx 2^{80}/ 2^{50}$ (high strength), respectively.

\begin{table}[!ht]
\centering
\caption{Example parameter sizes for the cognitive scheme.}
\label{table:paramcog}
\begin{tabular}{c|c|c|c|c|c|c}
$(d, k, l, n)$ & $m_{\text{it}}$ & $p_{\text{RG}}$ & BF & MitM & CH & GE \\
\hline\hline
$(5, 5, 24, 60)$ & 11 & 0.255 & $2^{22}$ & $2^{12}$ & $2^{11}$ & $\text{poly}(n)$\\
Samples required & - & 0 & 11 & 11 & 23 & 300 \\
\hline 
$(5, 10, 30, 130)$ &  24 & 0.252 & $2^{48}$ & $2^{28}$ & $2^{33}$ & $\text{poly}(n)$\\
Samples required & - & 0 & 24 & 24 & 24 & 650 \\
\hline 
$(5, 14, 30, 180)$ & 34 & 0.256 & $2^{68}$ & $2^{40}$ & $2^{40}$ & $\text{poly}(n)$\\
Samples required & - & 0 & 34 & 34 & 94 & 900 \\
\hline
$(5, 18, 30, 225)$ & 44 & 0.254 & $2^{87}$ & $2^{51}$ & $2^{51}$ & $\text{poly}(n)$\\
Samples required & - & 0 & 44 & 44 & 168 & 1125

\end{tabular}
\end{table}

\section{The Behavioural Biometric Scheme}
\label{sec:biometrics}
Our behavioural biometric authentication scheme is based on touch gestures with the assumption that a user exhibits similar behaviour when interacting with the touch screens of devices (such as smartphones, tablets and laptops) while remaining different from other users~\cite{touchalytics,unobserve-ndss}. In our scheme, we gather user touch gestures via a set of symbols $\Omega$. As mentioned before, the $(d + 1)$-classifier $D$ when given as input the template of a target user $\user$, should be able to decide two things: (a) whether the rendering of a symbol corresponds to a symbol in $\Omega$ or not, (b) whether the symbol matches the symbol template of $\user$. For each symbol in $\Omega$, the true positive rate (TPR) of $D$ is defined as the rate at which it correctly matches $\user$'s renderings of the symbol to $\user$'s template for that symbol. Similarly, the false positive rate (FPR) of $D$ is the rate at which it erroneously decides that $\adversary$'s rendering of the symbol matches $\user$'s template. The accuracy of $D$ for a symbol in $\Omega$ is defined as $\frac{1}{2}(\text{TPR} + 1 - \text{FPR})$. High accuracy therefore corresponds to high TPR and low FPR. For our purpose, we set TPR of $D$ close to $1.0$ for each symbol, and try to minimize the FPR. 

\subsection{Choice of Symbols}
\label{sub:symbols}
Previous research on gesture-based biometric authentication on touch screen devices indicate that the swipe gesture has good distinguishing characteristics~\cite{touchalytics, unobserve-ndss}. However, to obtain good accuracy, a large number of swipes need to be gathered~\cite{unobserve-ndss, chauhan2016gesture}. Our first criterion for designing symbols is that they should be rich enough as to simulate multiple swipes. The second property we require is that the symbol should be hard for $\adversary$ to mimic even after observing $\user$ render the symbol. The third consideration is that the symbol should be easily repeatable by $\user$ even when there is a gap between successive authentications. Finally, the symbols should be distinct from each other so that the classifier $D$ can easily distinguish between $d$ different symbols. With these properties in mind, we chose four different sets of symbols: \emph{easy words}, \emph{complex words}, \emph{easy figures} and \emph{complex figures}, shown in Table~\ref{tab:symbols} together with the mapping to elements of $\mathbb{Z}_5$. Some details follow:

\begin{itemize}
\item \emph{easy words:} This set corresponds to English words for the numbers, and serves as a base 
\iffulledition
case since the mapping is straightforward. However, these words can potentially be prone to mimicking. 
\else
case.
\fi
\item \emph{complex words:} Previous research shows that 
\iffulledition
some letters in the English alphabet are more difficult to write cursively than others because they contain more number of turns~\cite{kao1983control, van2011review}. The letters \emph{b, f, g, h, k, m, n, q, t, u, w, x, y,} and \emph{z} provide more turns than others~\cite{kao1983control}. 
\else
the letters \emph{b, f, g, h, k, m, n, q, t, u, w, x, y,} and \emph{z} are more difficult to write cursively than others because they contain more number of turns~\cite{kao1983control, van2011review}.
\fi
Our hypothesis was that if these letters are difficult to write, they might show more variation across users and also be difficult to mimic. Our user study shows the plausibility of our hypothesis, as complex words were the most resilient among all symbol sets against observation attacks. 

\iffulledition
We constructed five words from these 14 letters with at most one letter from the 14 in all words. The length of the words was fixed at four, since it was found through our implementation on a smartphone (see Section~\ref{sec:implement}) that words with more than 4 letters were difficult for users to write on the touch screen. 
\else
We constructed five words from these 14 letters each of length 4, since our smartphone implementation (see Section~\ref{sec:implement}) showed that higher length words were difficult to confine in the screen.
\fi
As it is difficult to construct meaningful words without vowels, we allowed one vowel to be present in each word. For both sets of words, we did not use capital letters as they show less variation among users~\cite{capital}.
\item \emph{easy figures:} This set contains numbers written in blackboard bold shape. We chose blackboard bold style to allow for more richness in the symbol. Another property of these figures is that the user can render them by starting at the top left most point and draw them without lifting the finger by traversing in a down and right manner. This removes a drawback present in the next symbol set, namely different segments of the symbol may be drawn at different order and direction each time, resulting in high variability within the user's drawings.
\item \emph{complex figures:} These figures were constructed by following some principles: no dots or taps \cite{unobserve-ndss, chauhan2016gesture}, the users finger must move in all directions while drawing the symbol and the symbol should have many sharp turns and angles \cite{Sherman}. These properties would potentially make the resulting figures harder to mimic. 
%The choice of the figures were also made so that there is a mnemonic association~\cite{mnemonic} with the number it maps to. 
\end{itemize}

The final choice of the symbol set was based on the results of our user study presented in Section~\ref{sec:results}.

%\begin{table}[!ht]
%	\begin{center}
%    \begin{tabular}{  m{0.8cm} | m{1cm} | m{1cm} | m{1cm} | m{1cm} | m{1cm}  }
%		%\hline
%		response & 0 & 1 & 2 & 3 & 4 \\
%		\hline\hline
%		easy words & $\mathtt{zero}$ & $\mathtt{one}$ & $\mathtt{two}$ & $\mathtt{three}$ & $\mathtt{four}$ \\
%		complex words & $\mathtt{xman}$ & $\mathtt{bmwz}$ & $\mathtt{quak}$ & $\mathtt{hurt}$ & $\mathtt{fogy}$ \\
%		easy figures & \includegraphics[width=0.7cm]{figures/new_00.png} & \includegraphics[width=0.8cm]{figures/new_11.png} & \includegraphics[width=0.7cm]{figures/new_22.png} &
%		 \includegraphics[width=0.7cm]{figures/new_33.png} & \includegraphics[width=0.6cm]{figures/new_44.png}\\
%		complex figures & \includegraphics[width=1.1cm]{figures/place00.png} & \includegraphics[width=1.1cm]{figures/place11.png} & \includegraphics[width=1.1cm]{figures/place22.png} &
%		 \includegraphics[width=1.1cm]{figures/place33.png} & \includegraphics[width=1.1cm]{figures/place44.png}
%    \end{tabular}
%    \caption{Mapping of responses ($d = 5$) to symbols}
%    \label{tab:symbols}
%    \end{center}
%\end{table}

\begin{table}[!ht]
	\begin{center}
    \begin{tabular}{  c | c | c | c | c | c  }
		%\hline
		response & 0 & 1 & 2 & 3 & 4 \\
		\hline\hline
		easy words & $\mathtt{zero}$ & $\mathtt{one}$ & $\mathtt{two}$ & $\mathtt{three}$ & $\mathtt{four}$ \\
		complex words & $\mathtt{xman}$ & $\mathtt{bmwz}$ & $\mathtt{quak}$ & $\mathtt{hurt}$ & $\mathtt{fogy}$ \\
		easy figures & \raisebox{-.5\height}{\includegraphics[width=0.5cm]{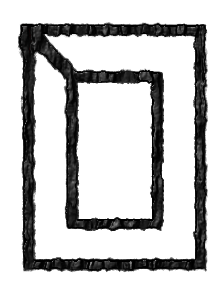}} & \raisebox{-.5\height}{\includegraphics[width=0.5cm]{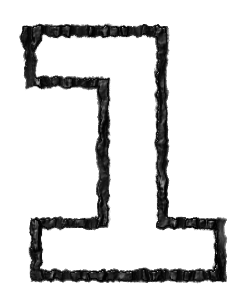}} & \raisebox{-.5\height}{\includegraphics[width=0.5cm]{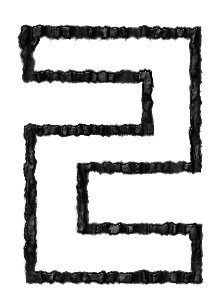}} &
		 \raisebox{-.5\height}{\includegraphics[width=0.5cm]{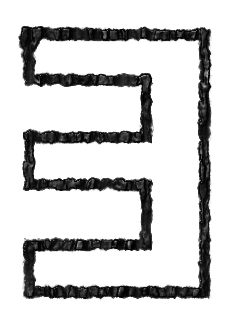}} & \raisebox{-.5\height}{\includegraphics[width=0.45cm]{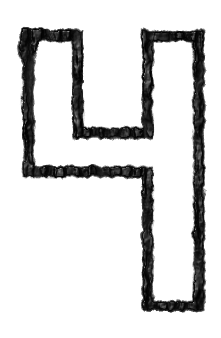}}\\
		complex figures &  \raisebox{-.5\height}{\includegraphics[width=0.7cm]{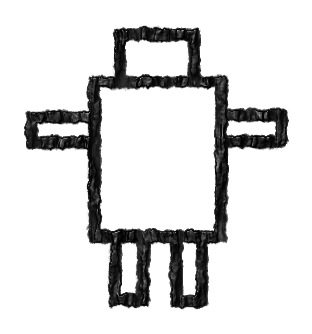}} & \raisebox{-.5\height}{\includegraphics[width=0.7cm]{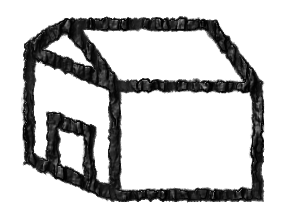}} & \raisebox{-.5\height}{\includegraphics[width=0.7cm]{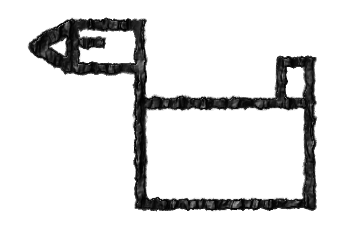}} &
		 \raisebox{-.5\height}{\includegraphics[width=0.7cm]{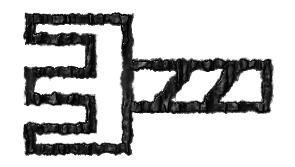}} & \raisebox{-.5\height}{\includegraphics[width=0.7cm]{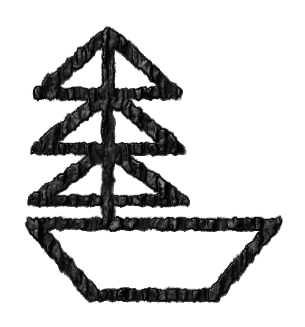}}
    \end{tabular}
    \caption{Mapping of responses ($d = 5$) to symbols.}
    \label{tab:symbols}
    \end{center}
\end{table}

\subsection{Choice of classifier}
Our classifier is based on the dynamic time warping (DTW) 
\iffulledition
algorithm~\cite{dtw-history, dtw-muller}. 
\else
algorithm~\cite{dtw-muller}. 
\fi
We picked DTW because: (a) all the chosen symbols naturally exhibit features that are a function of time, (b) DTW is claimed to be the best performing distance based classifier \cite{Ding}, and (c) DTW has been previously known to perform well with a small number of training samples (5-10) \cite{Nguyen} (which is important to minimize registration time). Given two time-series $q_1$ and $q_2$ of length $\tau_1$ and $\tau_2$, respectively, the algorithm creates a $\tau_1 \times \tau_2$ distance matrix where the $(i, j)$th element is the (squared) Euclidean distance $(q_1(i) - q_2(j))^{2}$, for $1 \le i \le \tau_1, 1 \le j \le \tau_2$. The output of the algorithm is the \emph{warped} path between the two time-series which minimizes the sum of distances of its constituent cells (satisfying some constraints such as starting at the bottom-left corner and finishing at the top-right corner). See~\cite{dtw-muller} for more details. For our purposes, we assume that the output $\text{DTW}(q_1, q_2)$ is the sum of distances (without caring about the actual path). We shall call this distance, the DTW distance between $q_1$, $q_2$. Assume that there is a set $Q$ of features, each of which is a time-series. Let $\hat{Q}$ represent the set of templates of the features in $Q$. We shall describe how we obtain such templates later. For now, assume that the template for each feature is a single time series. Given a test sample of these features (for authentication), also represented ${Q}$, the multi-dimensional DTW distance between $\hat{Q}$ and ${Q}$ is defined as~\cite{multi-dtw}:
\iffulledition
\[
\text{DTW}(\hat{Q}, Q) = \sum_{i = 1}^{|Q|} \text{DTW}(\hat{q}_i, q_i). 
\]
\else
$\text{DTW}(\hat{Q}, Q) = \sum_{i = 1}^{|Q|} \text{DTW}(\hat{q}_i, q_i)$.
\fi
We note that multi-dimensional DTW can also be defined by extending the Euclidean distance from one-dimension to $|Q|$ dimensions and then computing the DTW distance~\cite{multi-dtw}. However, our selected features originated from different sensors of the device (e.g., touch event or gyroscope), which have different sampling times and rates. We, therefore, chose the definition above so that we do not have any alignment issues and hence each dimension is treated independently. 

\subsection{Template Creation}
The set of templates $\hat{Q}$ is created as follows. For each user $\user$ and for each symbol in $\Omega$ we obtain 
$t$ sample renderings of the symbol. These samples result in $t$ time-series for each feature. Fix a feature. We 
take one of the $t$ time-series at a time, compute its DTW distance with each of the $t-1$ remaining time-series, 
and sum the distances. The time-series that gives the minimum cumulative sum is chosen as the \emph{optimal feature 
template}. The process is repeated for all features to create the template $\hat{Q}$. Note that, this means that the optimal feature templates in $\hat{Q}$ may not belong to the same sample. We in fact have two sets of optimal templates. The first template is used to check if $\user$ produced a valid rendering of a symbol from $\Omega$. For this template, we only use the $\mathtt{x}$, $\mathtt{y}$ coordinate time series features (see Table~\ref{table:features}). We denote this template by $\hat{Q}_{\text{sym}}$. The other template is used to check whether the rendering itself comes from the target user $\user$ or some one else. We shall simply denote this template by $\hat{Q}_{\text{user}}$. 

\subsection{Classification Decision}
Given a set of feature values $Q$ from a sample, the decision is made based on whether $\text{DTW}(\hat{Q}, Q)$ lies below a threshold. More specifically, the threshold is calculated as $\hbar \doteq \mu + z \sigma$, where $\mu$ is the mean DTW distance between the user's optimal template $\hat{Q}$ and all of the user's $t$ samples in the registration phase \cite{li2016whose}. Similarly $\sigma$ is the standard deviation. The parameter $z \ge 0$ is a global parameter that is set according to experimental data from a global set of users, and hence remains the same for every user. The notation $\hbar_{\text{sym}}$ denotes the threshold corresponding to $\hat{Q}_{\text{sym}}$, and $\hbar_{\text{user}}$ denotes the threshold corresponding to $\hat{Q}_{\text{user}}$. Our classification works as follows. 
\begin{itemize}
\item \textit{Step 1:} If for a given challenge $c = (a, w)$, $x \cap a \ne \emptyset$ (i.e., the non-empty case), $\server$ first gets the target symbol by computing $f$. Then, $\server$ computes the DTW distance between $\hat{Q}_{\text{sym}}$ and the sample. If the distance is greater than $\hbar_{\text{sym}}$ then $\user$ is rejected. Otherwise, $\server$ moves to Step 2. On the other hand, if it is the empty case, $\server$ computes the DTW distance between the incoming sample and $\hat{Q}_{\text{sym}}$ for each possible symbol. $\server$ assumes the symbol entered by $\user$ is the one with the least distance. After establishing the symbol, the distance is compared with $\hbar_{\text{sym}}$ for that symbol, and server accordingly rejects or goes to Step 2.
\item \textit{Step 2:}  In this step, the DTW distance between the sample and $\hat{Q}_{\text{user}}$ of the symbol is computed. If the distance is greater than $\hbar_{\text{user}}$ the user is rejected, otherwise it is accepted. 
\end{itemize}

\subsection{Feature Identification}
We identified three different categories of features: touch features, stylometric features and device-interaction features. Touch features correspond to how a user's finger interacts with the touch screen of the device.
\iffulledition
These include raw features obtained through the device's touch sensors as well as derived features (such as velocity, acceleration, and force). 
\fi
Stylometric features correspond to how a user renders a symbol on the touch screen. For example, the area covered by the symbol, and the angles between successive coordinate points in the symbol. 
\iffulledition
Stylometric features are constructed using the raw and derived features from the touch sensor data. Finally, device-interaction features originate from the way the user interacts with the device through a medium other than the touch screen, e.g., accelerometer and gyroscope. 

We identify 19 types of local features by exploring the literature \cite{xu-soups, chauhan2016gesture, bo2013silentsense, kinwrite} and obtain a feature vector of 40 dimensions, as shown in Table \ref{table:features}. 
\else
Finally, device-interaction features originate from the way the user interacts with the device through a medium other than the touch screen, e.g., accelerometer and gyroscope. We identify 19 types of local features by exploring the literature \cite{xu-soups, chauhan2016gesture, bo2013silentsense, kinwrite} and obtain a feature vector of 40 dimensions, as shown in Table \ref{table:features}. 
\fi
Each feature is a set of points arranged in time (time series) as the user renders a symbol on the touch screen of the device. The features shown in Table \ref{table:features} are mostly self explanatory. For the mathematical description of curvature, slope angle and path angle we refer the reader to~\cite{kinwrite}. Since different features have different range of values, we perform a standard $z$-score normalization on each 
\iffulledition
feature. 
\else
feature. Figure \ref{fig:axisxplot} shows the feature $\mathtt{x}$, i.e., the $x$-coordinate, as a time series for the complex word ``$\mathtt{xman}$.''  For ease of view, we show the feature without normalization. Observe that the way the word is written varies between different users (two samples from User 1, User 2 and User 3), while remains similar for the same user (User 1-A and 1-B).
\fi

\begin{table*} [!ht]
\centering
\caption{List of features.\label{table:features}}
%\caption{List of Features}
%\label{table:features}
\resizebox{1.0\textwidth}{!}{
\begin{tabular}{c|p{4.5cm} |c | c |c|c|c|c|c} 
 \# & Touch feature & Symbol  & \# & Stylometric feature & Symbol & \# & Device-interaction feature & Symbol\\
\hline\hline
 1. & Coordinates and change in coordinates & $\mathtt{x}$, $\mathtt{y}$ , $\delta\mathtt{x}$, $\delta\mathtt{y}$ & 8. & Top, bottom, left, right most point & $\mathtt{TMP}$, $\mathtt{BMP}$, $\mathtt{LMP}$, $\mathtt{RMP}$ & 15. & Rotational position of device in space & $\mathtt{R}_{\mathtt{x}}$, $\mathtt{R}_{\mathtt{y}}$, $\mathtt{R}_{\mathtt{z}}$  \\
	 2. & Velocity along coordinates & $\dot{\mathtt{x}}$, $\dot{\mathtt{y}}$    & 9. & Width: $\mathtt{RMP} - \mathtt{LMP}$, Height: $\mathtt{TMP} - \mathtt{BMP}$ & $\mathtt{width}$, $\mathtt{height}$& 16. & Rate of rotation of device in space & $\mathtt{G}_{\mathtt{x}}$, $\mathtt{G}_{\mathtt{y}}$, $\mathtt{G}_{\mathtt{z}}$  \\
	 3. & Acceleration along coordinates & $\ddot{\mathtt{x}}$, $\ddot{\mathtt{y}}$   &10. & Rectangular area: $\mathtt{width} \times \mathtt{height}$  & $\mathtt{area}$ & 17. & 3D acceleration force due to device's motion and gravity & $\mathtt{A}_{\mathtt{x}}$, $\mathtt{A}_{\mathtt{y}}$, $\mathtt{A}_{\mathtt{z}}$  \\
	 4. & Pressure and change in pressure & $\mathtt{p}$, $\delta \mathtt{p}$& 11. & Width to height ratio & $\mathtt{WHR}$ & 18. & 3D acceleration force solely due to gravity & $\mathtt{g}_{\mathtt{x}}$, $\mathtt{g}_{\mathtt{y}}$, $\mathtt{g}_{\mathtt{z}}$  \\
	 5. & Size and change in size & $\mathtt{s}$, $\delta \mathtt{s}$& 12. & Slope angle & $\theta_{\mathtt{slope}}$ & 19. & 3D acceleration force solely due to device's motion & $\mathtt{a}_{\mathtt{x}}$, $\mathtt{a}_{\mathtt{y}}$, $\mathtt{a}_{\mathtt{z}}$  \\
	 6. & Force:   $\mathtt{p} \times \mathtt{s}$ & $\mathtt{F}$ & 13. & Path angle & $\theta_{\mathtt{path}}$ &  & \\
	 7. & Action type:  finger lifted up, down or on touchscreen & $\mathtt{AT}$ & 14. & Curvature   & $\mathtt{curve}$ & & 
\end{tabular}
}
\end{table*}

%Since we are using symbols, global features, such as the central position and the average velocity, do not contain much useful information for distinguishing users~\cite{kinwrite} \textcolor{red}{this point is not clear.}. Instead 
\iffulledition
Figure \ref{fig:axisxplot} shows the feature $\mathtt{x}$, i.e., the $x$-coordinate, as a time series for the complex word ``$\mathtt{xman}$.''  For ease of view, we show the feature without normalization. Observe that the way the word is written varies between different users (two samples from User 1, User 2 and User 3), while remains similar for the same user (User 1-A and 1-B).
\fi
\iffulledition
\begin{figure}[!ht]
	\centering
    \includegraphics[width =9cm]{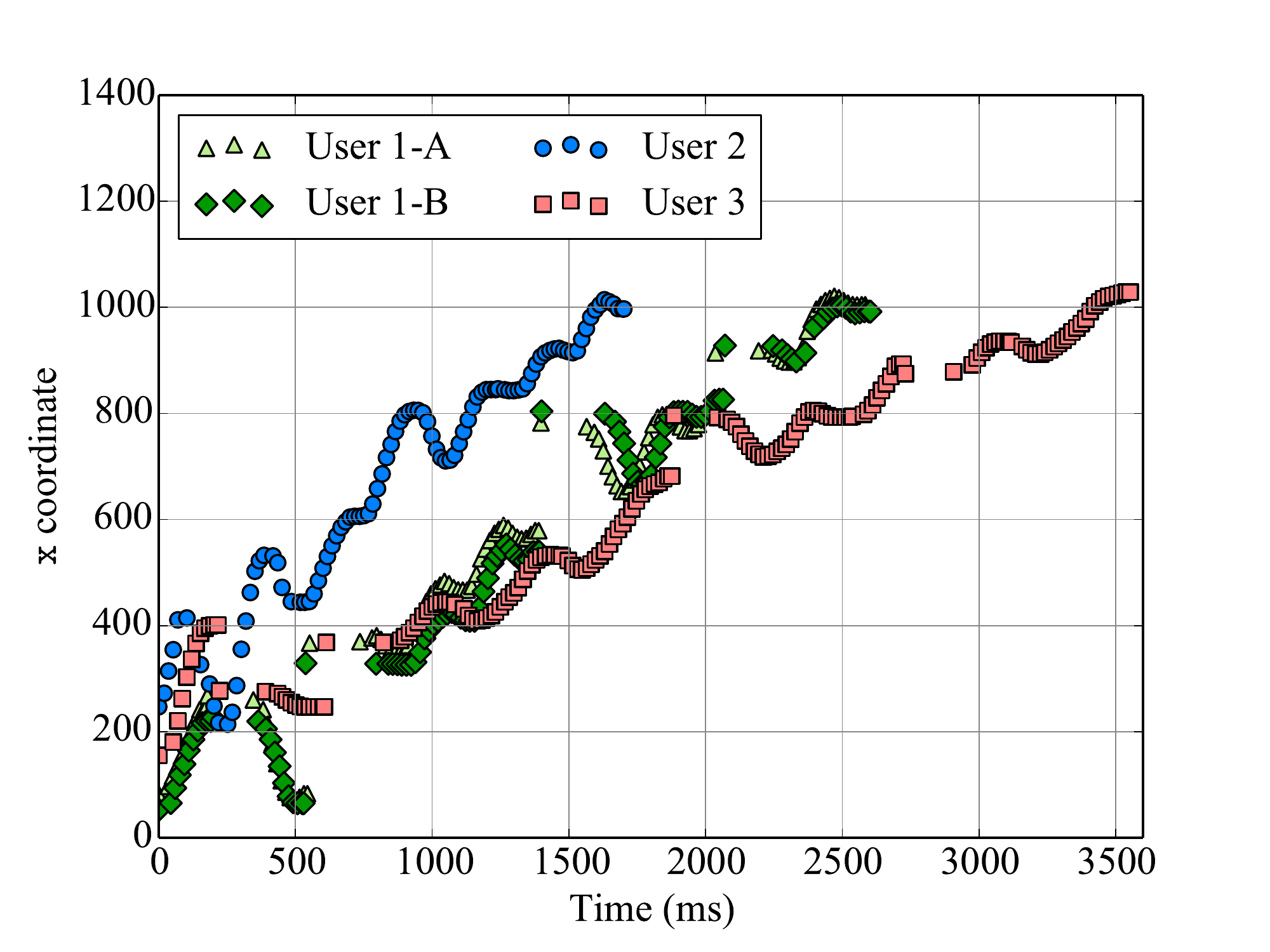}
  	\caption{Feature comparison of samples from three users.}
 	\label{fig:axisxplot}
\end{figure}
\else
\begin{figure}[!ht]
	\centering
    \includegraphics[width =5.3cm]{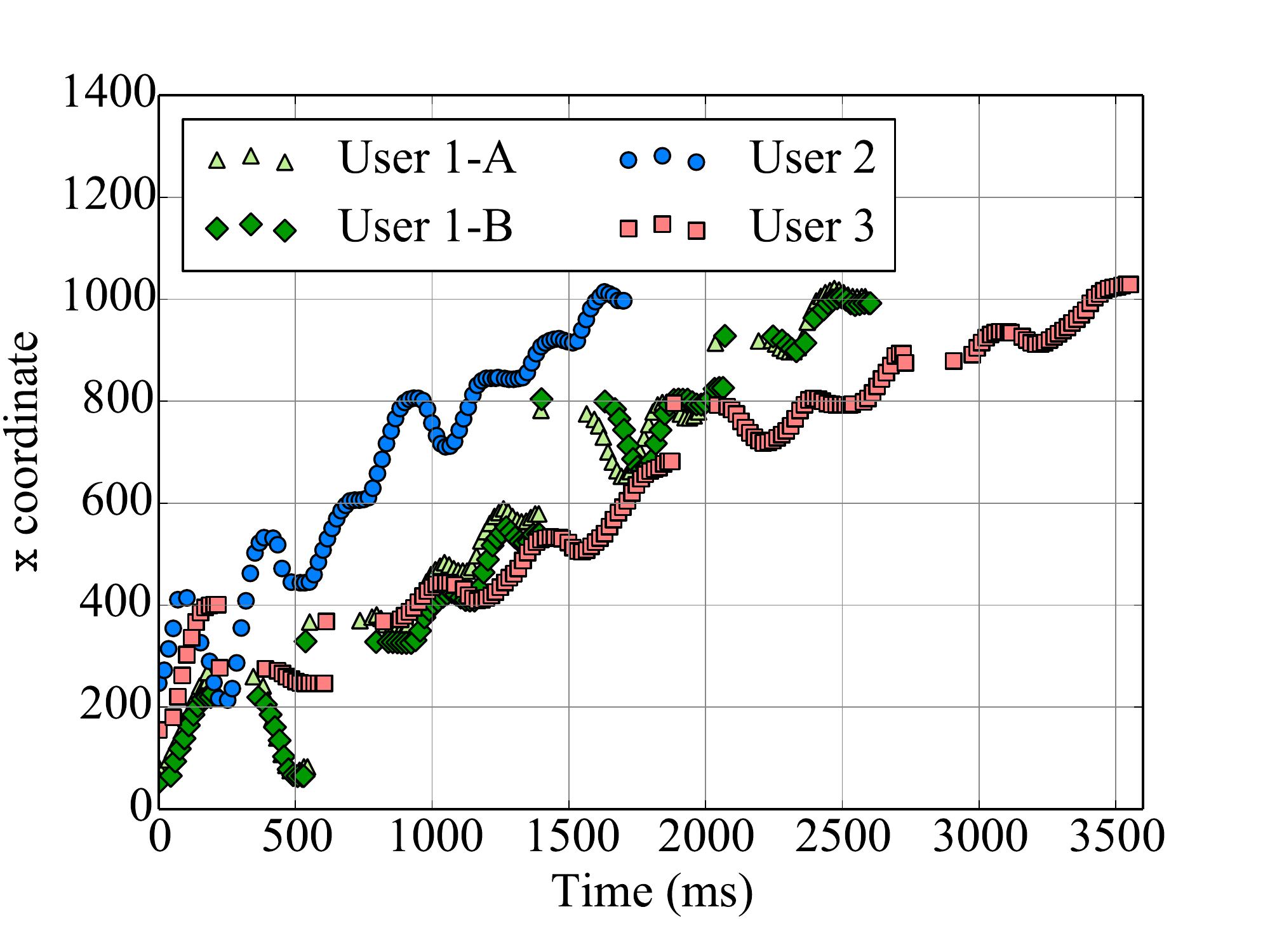}
  	\caption{Feature comparison of samples from three users.}
 	\label{fig:axisxplot}
\end{figure}
\fi

\subsection{Feature Selection}
In order to select the most distinguishing and non-redundant features from the list of 40 features for each symbol, we created our own variation of a wrapper method\footnote{i.e., feature subset selection is based on classification results.} based on sequential forward feature selection~\cite{sfs1, sfs2}. This is described in Algorithm~\ref{algo:feat-select}. The algorithm is given as input a total list of features $Q_{\text{tot}}$ and a symbol, and it outputs a selected list of features $Q$ for that symbol. The algorithm starts with an empty list and iteratively adds one feature at a time by keeping TPR = 1.0 and minimizing the FPR. The TPR and FPR values are calculated based on user-adversary pairs (see Section~\ref{sub:exp:phase1}). The algorithm terminates when all features in $Q_{\text{tot}}$ are exhausted. At the end, we are left with multiple candidate subsets for $Q$ (based on the number of features included). We pick the one which has  TPR = 1.0 and the least FPR as the final set of features. As we use a wrapper-based approach, running the classifier is an inherent part in feature selection. The algorithm therefore calls the Get $z$-List algorithm (Algorithm~\ref{algo:zlist}) as a subroutine (based on a similar procedure from~\cite{li2016whose}). This algorithm computes the $z$ values that give TPR of 1 and the least FPR for each possible feature subset.

\begin{algorithm}[!ht]
\iffulledition
\small
\else
\footnotesize
\fi
\SetAlgoLined
\SetAlCapSkip{1em}
\DontPrintSemicolon{}
\let\oldnl\nl% Store \nl in \oldnl
\newcommand{\nonl}{\renewcommand{\nl}{\let\nl\oldnl}}% Remove line number for one line
%\nonl\TitleOfAlgo{{Select Features}}
\SetKwInOut{Input}{input}
\Input{Set of all features $Q_{\text{tot}}$, a symbol $\in \Omega$, a set of user-attacker pairs $(\user, \adversary)$.}
Initialize $Q_{\text{sel}}^{(0)} \leftarrow \emptyset$, $i \leftarrow 0$.\;
\For {$j = 1$ to $|Q_{\text{\emph{tot}}}|$}{
	Set $i \leftarrow i + 1$.\;
	Create temporary feature subsets $Q_{j}$ by adding feature $q_j \in Q_{\text{tot}}$ to $Q_{\text{sel}}^{(i-1)}$.\;
	\For{each $(\user, \adversary)$ pair}{
		Run Get $z$-List algorithm (Algorithm~\ref{algo:zlist}) with inputs $Q_{j}$, $\user$ and $\adversary$ to get a $z$-list.\;
	}
	Sum TPR and FPR values for all users for each value of $0 \le z \le z_{\max}$ in the $z$-list. 
}
Let $Q_j$ be the temporary feature subset that has the minimum FPR sum with TPR sum equal to 1.0.\; 
Set $Q_{\text{sel}}^{(i)} \leftarrow Q_{\text{sel}}^{(i-1)} \cup \{q_j\}$, $Q_{\text{tot}} \leftarrow Q_{\text{tot}} - \{q_j\}$.\;
Repeat Steps 2-9 until $Q_{\text{tot}}$ is empty.\;
Return $Q \doteq Q_{\text{sel}}$ from the $Q_{\text{sel}}^{(i)}$'s that has the least $\text{FPR}$.\;
\caption{Select Features}
\label{algo:feat-select}
\end{algorithm}

\begin{algorithm}[!th]
%\SetAlgorithmName{Algorithm}{List of subroutines} 
\iffulledition
\small
\else
\footnotesize
\fi
\SetAlgoLined
\SetAlCapSkip{1em}
\DontPrintSemicolon{}
\let\oldnl\nl% Store \nl in \oldnl
\newcommand{\nonl}{\renewcommand{\nl}{\let\nl\oldnl}}% Remove line number for one line
%\nonl\TitleOfAlgo{{Get $z$-List}}
\SetKwInOut{Input}{input}
\Input{Feature subset $Q$, registration and test samples from $\user$, test samples from $\adversary$.}
For the features in $Q$, find the optimal template $\hat{Q}$ together with $\mu$ and $\sigma$ from $\user$'s registration samples.\;
Initialize an empty $z$-list.\;
Initialize $z \leftarrow 0$, $\text{step} \leftarrow 0.125$, $\text{TP} \leftarrow 0$, $\text{FP} \leftarrow 0$.\;
\While{$z \le z_{\max} \doteq 10$}{
	Set $\hbar \leftarrow \mu + z\sigma$\;
	\For{each test sample from $\user$}{
		\If{\emph{DTW} distance between $\hat{Q}$ and test sample is $\le \hbar$}{
			$\text{TP} \leftarrow \text{TP} + 1$.\;
		}
	}
	\For{each test sample from $\adversary$}{
		\If{\emph{DTW} distance between $\hat{Q}$ and test sample is $\le \hbar$}{
			$\text{FP} \leftarrow \text{FP} + 1$.\;
		}
	}
	Compute TPR and FPR by normalizing the TP and FP values.\;
	Update $z$-list with the tuple $(z, \text{TPR}, \text{FPR})$.\;
	Set $z \leftarrow z + \text{step}$.\; 
}
Return $z$-list.\;
\caption{Get $z$-List}
\label{algo:zlist}
\end{algorithm}

\section{Implementation}
\label{sec:implement}
We implemented the BehavioCog scheme as an app for Android based smartphones. We used a set of emojis, called \emph{twemojis},\footnote{Copyright 2016 Twitter, Inc and other contributors, \url{https://github.com/twitter/twemoji}} as the global set of objects for the cognitive scheme. The set consisted of $n = 180$ emojis. While the system is tunable, we used the parameters $(k, l, n) = (14, 30, 180)$ for our user study (corresponding to the medium strength adversary defined in Section~\ref{sub:paramcog}). Figure~\ref{fig:logina} shows an example challenge. For the behavioural biometric scheme, we used the implementation of DTW from the FastDTW library \cite{Salvador} and used a radius of 20. The radius controls how much the optimal path can drift from the diagonal of the two dimensional matrix representing the two time series to be aligned, and helps in finding more accurate shortest paths between the two. Raw touch features such as coordinates, pressure, size, and action type are extracted using the standard Android API \cite{touchapi} from the touch sensor of the device. The Android Motion Sensor API \cite{motionapi} is used to extract the features for device-interaction features. Our implementation handles any set of symbols used for $\Omega$. However, through our user study we found the set of complex words to be best in terms of repeatability and hardness of mimicking (see Section~\ref{sec:results}). Figure~\ref{fig:loginb} shows an example response entered by the user, using the complex word $\mathtt{fogy}$. Note that we use a dotted trace on the screen. Showing the full trace makes it easier for an attacker to observe fine details of a target user's symbols, while showing no trace at all makes it hard for the user since they cannot see what they already drew or wrote. The dotted trace is a compromise. 

\iffulledition
\begin{figure}[ht]
\centering
\subfloat[challenge \label{fig:logina}]{%
      \includegraphics[width=0.14\textwidth]{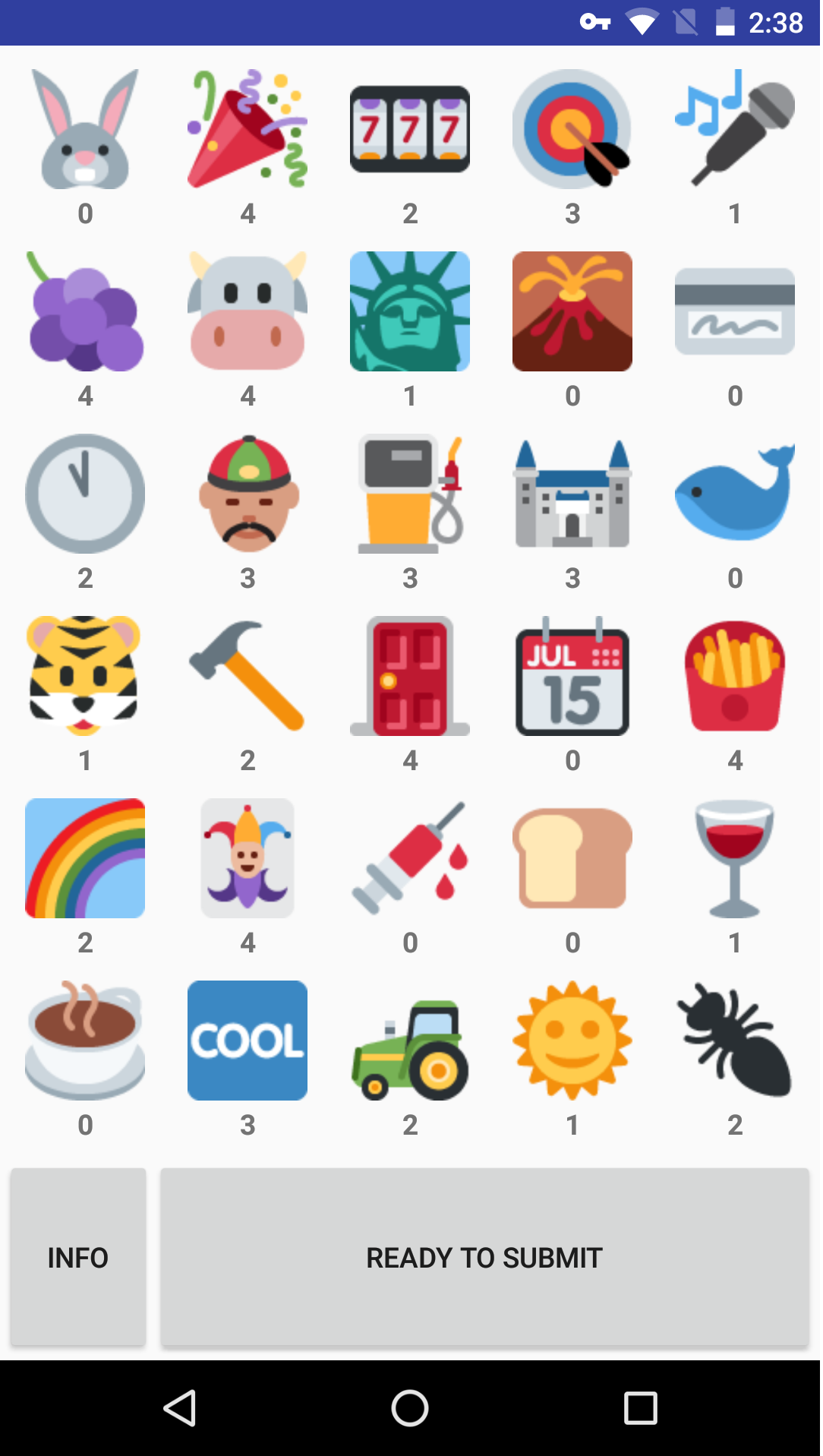}
}
%\hfill
\subfloat[response \label{fig:loginb}]{%
      \includegraphics[width=0.14\textwidth]{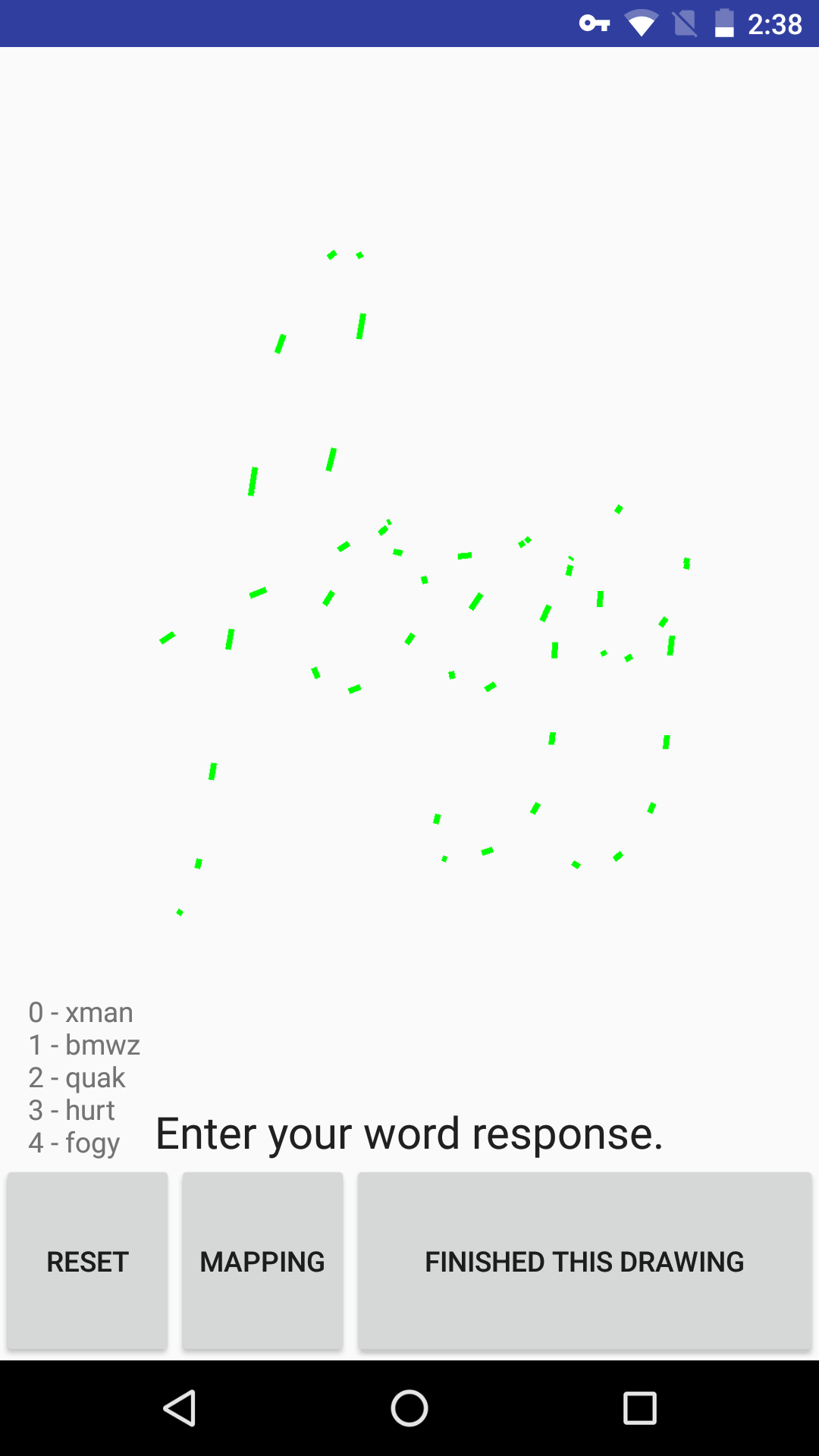}
}
\caption{An example challenge and response in our implementation of BehavioCog.}
\end{figure}
\else
\begin{figure}[ht]
\centering
\subfloat[challenge \label{fig:logina}]{%
      \includegraphics[width=0.14\textwidth]{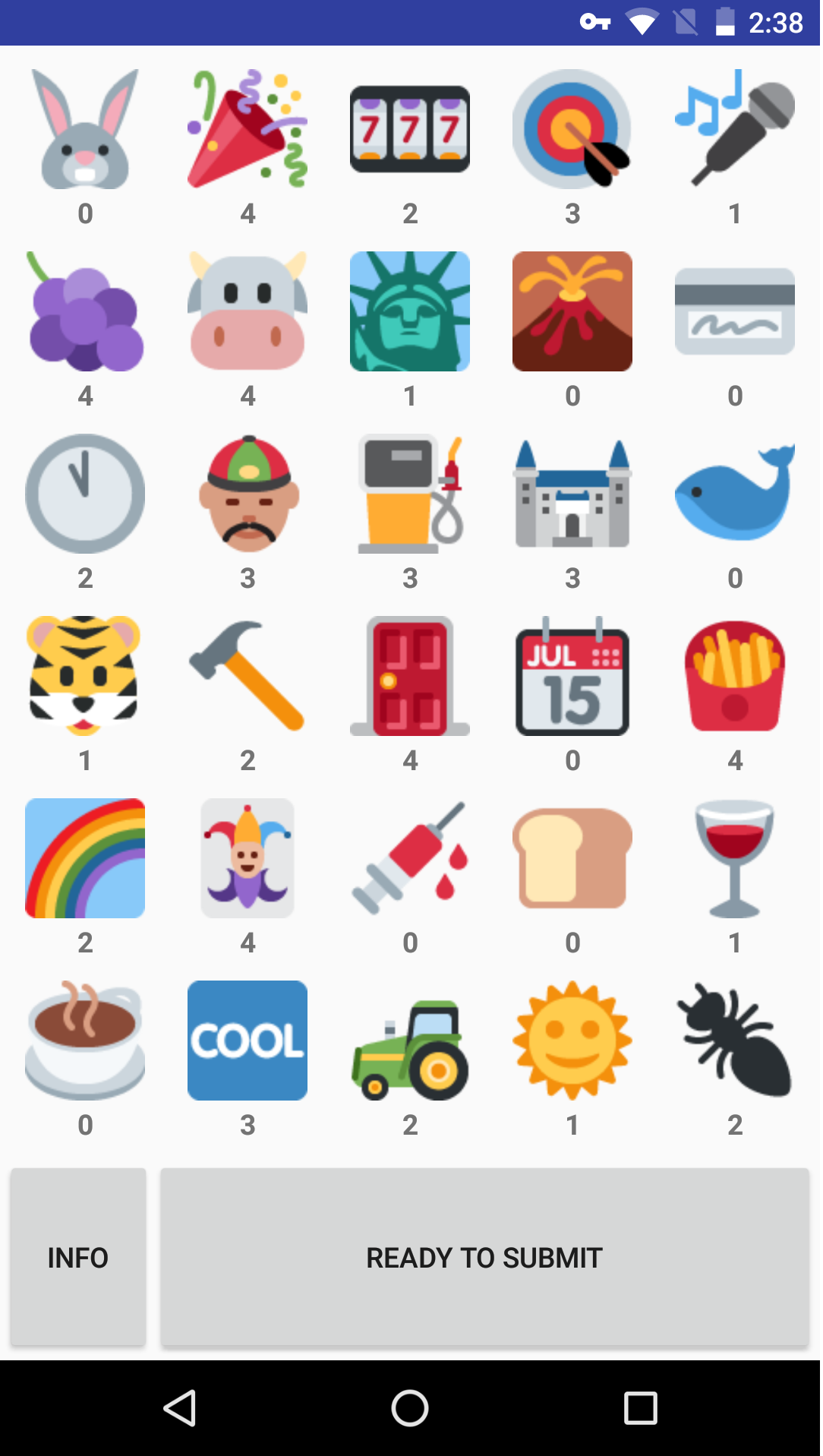}
}
%\hfill
\subfloat[response \label{fig:loginb}]{%
      \includegraphics[width=0.14\textwidth]{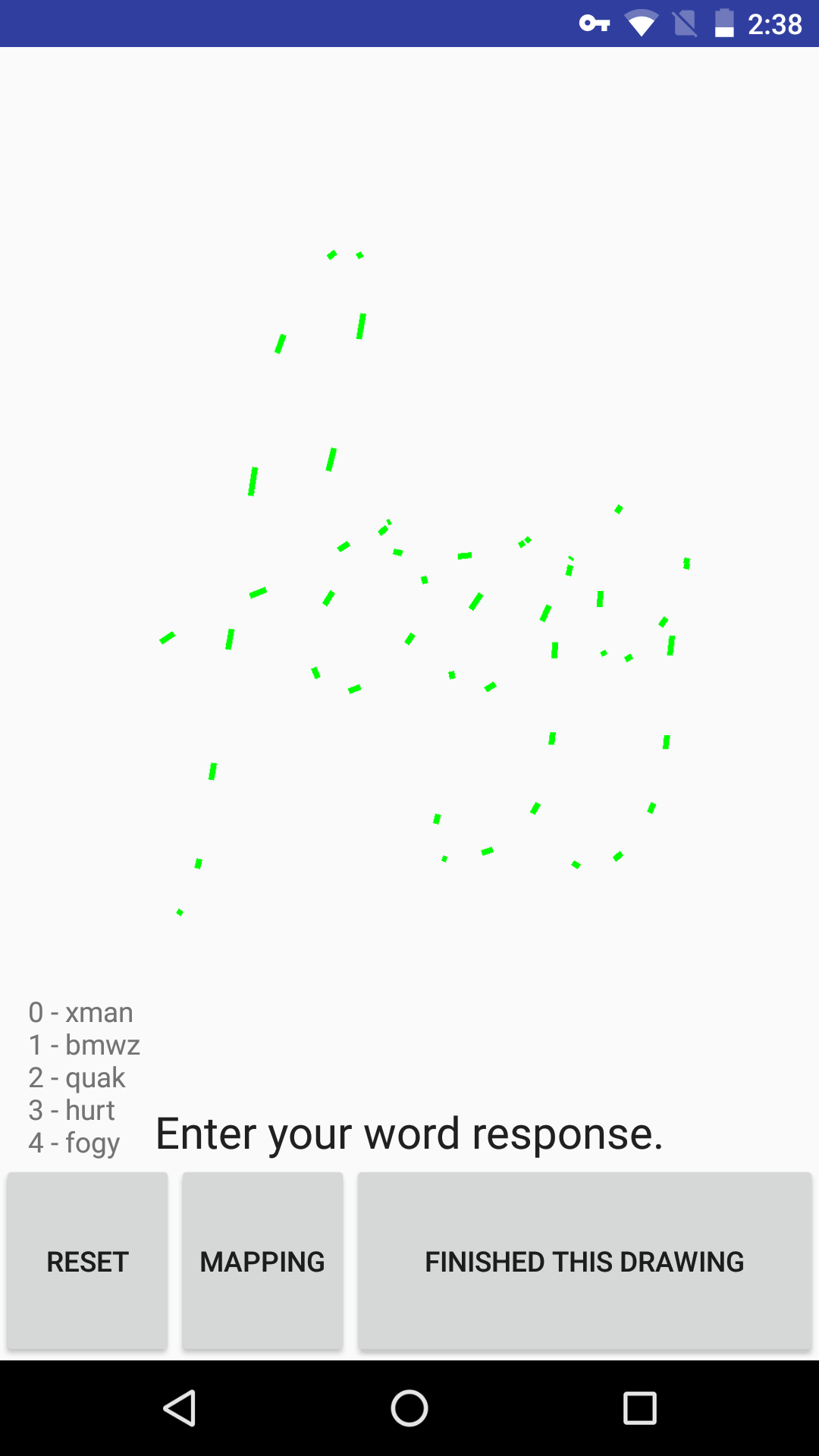}
}
\caption{An example challenge and response in our implementation of BehavioCog.}
\end{figure}
\fi

Also note that we explicitly show the mapping of responses to complex words. While this is fine, we were also interested in knowing whether users could be trained to remember the mapping. To do so, we created a few mnemonic helpers for users to remember the mapping. These are shown in Table~\ref{table:mnemonic}. The mnemonic strategy used is a mixture of the (rhyming) peg method, keyword method and picture-based mnemonics~\cite{peg-mnemonics, blocki}. 
\iffulledition
We reiterate that the map $\text{sym}(r)$, for $r \in \mathbb{Z}_d$ does not need to be a secret, and therefore the user is not required to remember the mapping as far as security is concerned. 
\fi
\newcommand{\mysize}{0.8cm}
 \begin{table}[!ht]
	\begin{center}
    \begin{tabular}{  r | c | l  }
		
		$r$ & Mnemonic & Word \\ \hline\hline
				 0 & hero &  $\mathtt{xman}$ is our hero\\
		
				1 & run & $\mathtt{bmwz}$ runs on the street\\
		
				2 & \raisebox{-.5\height}{\includegraphics[width=\mysize]{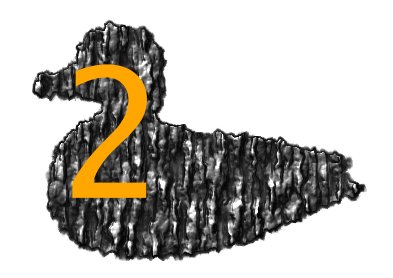}} & the duck goes $\mathtt{quak}$\\
		
				3 & \raisebox{-.5\height}{\includegraphics[width=\mysize]{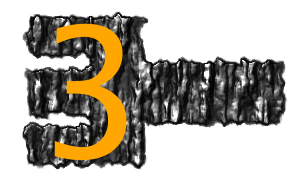}}& I got $\mathtt{hurt}$ by a trident\\
		
				4 & \raisebox{-.5\height}{\includegraphics[width=\mysize]{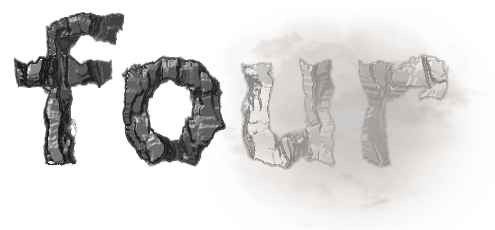}}& can't see four when it's $\mathtt{fogy}$

    \end{tabular}
    \caption{Mnemonic mapping of cognitive response to complex words.}
    \label{table:mnemonic}
    \end{center}
\end{table}

%\begin{table}[!ht]
%	\begin{center}
%    \begin{tabular}{  l | m{1.5cm} | m{4cm}  }
%		
%		response & mnemonic & word \\ \hline\hline
%				 0 & hero &  \texttt{xman} is our hero\\
%		
%				1 & run & \texttt{bmwz} runs on the street\\
%		
%				2 &\includegraphics[width=1.4cm]{figures/duck0.png} & the duck goes \texttt{quak}\\
%		
%				3 &\includegraphics[width=1.4cm]{figures/trident0.png}& I got \texttt{hurt} by a trident\\
%		
%				4 &\includegraphics[width=1.5cm]{figures/fogy00.png}& can't see four when it's \texttt{fogy}
%			
%
%
%    \end{tabular}
%    \caption{Mnemonic mapping of cognitive response to complex words.}
%    \label{table:mnemonic}
%    \end{center}
%\end{table}

Based on the example parameter sizes for the cognitive scheme (Table~\ref{table:paramcog}) and the results from the user study, we recommend the parameters shown in Table~\ref{table:paramfull}. The columns labelled ``Sessions'' indicate whether the target is a medium-strength or high-strength adversary $\mathcal{A}$ as discussed in Section~\ref{sub:paramcog}. The acronym CW stands for complex words. Based on our experiments, these words gave the best average FPR. The security column shows the probability the $\adversary$ can succeed in impersonating the user by randomly guessing the response and mimicking the corresponding behavioural biometric symbol. For complex words, the average FPR, denoted $\overline{\text{FPR}}$ was $0.05$ (see Table~\ref{table:phase1results1}). By setting $p_{\text{RG}} = 0.25$ (approximately the random guess probability for the cognitive scheme) and multiplying the two quantities we estimate the total impersonation probability of $\adversary$. For reference, the same probability for a $4$-digit PIN is $1 \times 10^{-4}$, and for a $6$-digit PIN is $1 \times 10^{-6}$ (but with no security under observation).
\begin{table}[!ht]
\centering
\caption{Example parameter sizes for BehavioCog.}
\label{table:paramfull}
\begin{tabular}{c|c|c|c|c|c}
\multirow{ 2}{*}{$(d, k, l, n)$} & \multirow{ 2}{*}{$\gamma$} & Sessions & Sessions & \multirow{ 2}{*}{$\Omega$} & \multirow{ 2}{*}{Security} \\
& & (med. $\adversary$) & (high $\adversary$) & &\\
\hline\hline
$(5, 5, 24, 60)$ & 1 & 10 & 10 & CW & $1.3 \times 10^{-2}$ \\
$(5, 5, 24, 60)$ & 2 & 5 & 5 & CW & $1.5 \times 10^{-4}$\\
$(5, 5, 24, 60)$ & 3 & 3 & 3 & CW & $2 \times 10^{-6}$\\
\hline 
$(5, 10, 30, 130)$ & 1 & 24 & 24 & CW & $1.3 \times 10^{-2}$\\
$(5, 10, 30, 130)$ & 2 & 12 & 12 & CW & $1.5 \times 10^{-4}$\\
$(5, 10, 30, 130)$ & 3 & 8 & 8 & CW & $2 \times 10^{-6}$\\
\hline 
$(5, 14, 30, 180)$ & 1 & 94 & 34 & CW & $1.3 \times 10^{-2}$\\
$(5, 14, 30, 180)$ & 2 & 47 & 17 & CW & $1.5 \times 10^{-4}$\\
$(5, 14, 30, 180)$ & 3 & 31 & 11& CW & $2 \times 10^{-6}$\\
\hline 
$(5, 18, 30, 225)$ & 1 & 511 & 168 & CW & $1.3 \times 10^{-2}$\\
$(5, 18, 30, 225)$ & 2 & 255 & 84 & CW & $1.5 \times 10^{-4}$\\
$(5, 18, 30, 225)$ & 3 & 170 & 56& CW & $2 \times 10^{-6}$\\
\end{tabular}
\end{table}

\section{User Study}
\label{sec:experiments}
In this section we give details of our user study. The results from the study are shown in Section~\ref{sec:results}. An Android based Nexus 5x smartphone was used in the experiments in a controlled setting where the user sits on a chair and performs the experiments. We got the ethics approval for our user study. The total number of participants in the study was 41. Our experiments were conducted in three phases. In the first phase, we only collected touch biometric data from the users to select the best features for each symbol in terms of repeatability and hardness of mimicking using our feature selection algorithm, and to determine which symbol set performs the best. In the second phase, the complete hybrid authentication scheme was tested to assess the usability and security of the scheme. In the third phase, we performed extended cognitive training and testing to check if more careful user training can reduce cognitive errors present in Phase 2. 

\subsection{Phase 1}
\label{sub:exp:phase1}
This phase involved collecting biometric samples from 22 (actual) participants: 8 females and 14 males for different symbol sets.  As some users agreed to provide biometric samples for more than one symbol set, we ended up with 40 (logical) users. The 40 users were then divided equally into four groups, one for each symbol set. The phase involved two sessions. In the first session, the user was asked to render each of the five symbols 13 times on the smartphone screen. The first 10 were used to construct templates, and the last 3 were reserved for testing. The whole session was video recorded. 

The second session started a week after the first session. Each user was asked to render all the symbols given to them in the first session a further three times. These three samples were then used to test the repeatability of symbols from the same user. Each user was also asked to act as an attacker for a particular target user and vice versa from the same group (same symbol set). Thus, it was ensured that two participants from the same group perform observation attacks against each other. The (pretend) attacker watched a video of how a target user writes a particular symbol. The attacker was given unrestricted control to the playback of the video and was allowed to take notes. The attacker was then asked to mimic the target user by writing each symbol three times. 
 \subsection{Phase 2}
For Phase 2, we fixed the set of symbols to complex words based on our results (described in Section~\ref{sec:results}). This phase had a total of 30 (actual) participants, 11 of whom had also participated in Phase 1. The 30 subjects included 11 females and 19 males with 7 users in the age group 21-25, 8 users in the age group 26-30, 13 in the age group 31-35 and 2 users over the age of 40. This phase also consisted of two sessions. The first session involved (cognitive and biometric) registration and authentication. The biometric registration and authentication were video recorded. The second session involved authentication, performing a random and video based observation attack against a target user, and filling a questionnaire. 

The 30 users were divided equally in three groups: Group 1, 2 and 3. The users in each group differed from other groups in the amount of time they were allowed to do registration. In all the groups, user chose 14 emojis as their pass-emojis from the pool of 180 emojis. For Group 1, we further asked the users to write each of the five complex words three times to create their templates. %In our user studies, the average taken time to complete the registration process for \textit{Group1} users is x seconds.
 %and has to remember the mnemonic mappings between cognitive responses and the symbolsand remembering symbols while doing the verification process.
The registration for Group 2 included an extended training game to help them recognize their pass-emojis for better authentication accuracy and to help them familiarize with BehavioCog. The training game was divided into multiple steps in increasing order of difficulty, outlined below.

\iffulledition
\begin{enumerate}
\item The user was shown a fixed number of emojis with no assigned weights. This was initially set to 5. The screen contained at least one (random) pass-emoji and all others were random decoys. The user was told exactly how many of their pass-emojis were present and was asked to tap on them. This process was repeated by increasing the number of emojis shown from 5 to 25 in steps of 5 (with a corresponding increase in the number of pass-emojis).
\item In order to aid the user in associating responses in $\mathbb{Z}_5$ to the corresponding words, mnemonic associations were shown to the user as shown in Table~\ref{table:mnemonic}. The user was then given a series of easy questions with the correct answer being the complex word to be entered. An example question was: ``0 rhymes with hero, who is our hero? $\mathtt{xman}$ or batman?'' 
%and ``1 rhymes with run,  which car runs? \texttt{bmwz} or audi?'' 
There were three different questions for each word, meaning the user wrote each word three times.
\item This step was the same as Step 1 except that (a) the user was not told how many of their pass-images were present, and (b) the number of images was increased from 5 to 30 in steps of 5. 
\item This step was the same as Step 3 except that (a) the images also had weights in $\mathbb{Z}_5$, and (b) the user had to compute $f$, map the response to the word and press one of five buttons corresponding to the correct word. 
\item This step was the same as Step 2 except that the questions asked were slightly more difficult, e.g., ``0 rhymes with hero, \_ \_ \_ \_ is our hero.'' The user had to write each symbol two more times.
\end{enumerate}
\else
\begin{enumerate}
\item The user was shown a fixed number (initially 5) of emojis without weights. At least one of  them was a (random) pass-emoji. The user was told exactly how many of their pass-emojis were present and was asked to tap on them. Gradually, the emojis were increased from 5 to 25 in steps of 5 (with a corresponding increase in pass-emojis).
\item To help the user associate responses in $\mathbb{Z}_5$ to words, mnmonic associations were shown to the user as shown in Table~\ref{table:mnemonic}. The user was then given a series of easy questions with the correct answer being the complex word. An example question was: ``0 rhymes with hero, who is our hero? $\mathtt{xman}$ or batman?'' 
%and ``1 rhymes with run,  which car runs? \texttt{bmwz} or audi?'' 
There were three different questions for each word, meaning the user wrote each word three times.
\item This step was the same as Step 1 except that (a) the user was not told how many of their pass-images were present, and (b) the number of images was increased from 5 to 30 in steps of 5. 
\item This step was the same as Step 3 except that (a) the images also had weights in $\mathbb{Z}_5$, and (b) the user had to compute $f$, map the response to the word and press one of five buttons corresponding to the correct word. 
\item This step was the same as Step 2 except that the questions asked were slightly more difficult, e.g., ``0 rhymes with hero, \_ \_ \_ \_ is our hero.'' The user had to write each symbol two more times.
\end{enumerate}
\fi

The users in Group 3 had the same registration process as Group 2 but with the number of iterations increased. Namely, 10 biometric samples of each word were collected (five each from Steps 2 and 5). Also, the number of iterations in Steps 1, 3, and 4 were twice more than those for Group 2. Immediately after registration, users from all groups were asked to attempt authentication of the complete authentication scheme. Each user was allowed 6 authentication attempts. These were marked as authentication attempts of Session 1. A same number of authentication attempts were repeated after a gap of one week (but with no registration this time) for Session 2. The users were given a couple of trial runs to get a feel of how the system works before their attempts were logged. 

To simulate attacks, each (attacking) user was assigned a target user. The attacker was asked to perform a random attack and a video based observation attack, and then attempt to impersonate the user three times each. 
\iffulledition
In the random attack, the attacker was asked to guess the cognitive response and write the corresponding word on the touch screen. In the video based observation attack, the attacker was allowed to watch authentication attempts of the target user as many times as they wanted.  Afterwards, the attacker was asked to impersonate the target user. To have more advantage in impersonation, an attacker needs to watch as many \textit{correct} authentication attempts of the target user as possible. Hence, we picked those users as target users who successfully authenticated at least 5 out of 6 times during Session 1. 

During Session 2, we also asked the user to pick their 14 pass-emojis by showing them the whole pool of emojis. This helps us to know how many emojis a user can recognize after a week. We also asked the users to pick 14 pass-emojis, which they believed belonged to their target (attacked) user. This helps us to know how many emojis an attacker is able to guess correctly for any target user. Finally, at the end of Session 2 users were asked to fill a questionnaire including a section to fill their demographics: sex, age group and handedness. The answers to the questions were scaled on a Likert 5 scale.  The questions in general asked the users about their perception of the usability and security of the proposed authentication scheme.  
\else
During Session 2, we also asked the user to pick their 14 pass-emojis by showing them the whole pool of emojis. We also asked the users to pick 14 pass-emojis, which they believed belonged to their target (attacked) user. Finally, at the end of Session 2 users were asked to fill a questionnaire asking users about their perception of the usability and security of our scheme.  
\fi
 
%The main goal of Phase 2 of the user study was to find out how our proposed authentication scheme performs in terms of repeatability and security against observation attacks. To measure repeatability, we checked how many times a user could get access to the system in the second session. The security of the system was measured by checking how many times the attacker got unauthorized access to the system. The usability of the scheme was measured by measuring the authentication time for each round of the scheme. We also measured how many times users could not access the system and the underlying reason: cognitive error or biometrics error.
 
\subsection{Phase 3}
This phase was carried out as an after thought. We observed a higher number of authentication errors in Session 2 as compared to Session 1 in Phase 2 of the study (e.g., users in Group 3 made 41\% cognitive errors in Phase 2 versus 15\% in Phase 1; see Table~\ref{table:phase2loginresults}.). 
\iffulledition
These increased errors could either be due to users' failure to recognize pass-emojis after a gap of one week or due to error in computing $f$. 
Consequently, we decided to carry out a third phase of the user study to find out the root cause and decrease the error rate. 
\else
To find the cause of errors, we carried out Phase 3 of the user study.
\fi
The participants in this phase were only the users from Group 3, since they received the maximum amount of training and were most familiar with the authentication scheme. Phase 3 again consisted of 2 sessions. In the first session, each user was given an extended cognitive training immediately followed by authentication attempts. Session 2 happened a week after Session 1 and only involved authentication attempts. In the extended cognitive training, each user was shown their 14 pass-emojis one by one for 10 seconds followed by a 3 second cool off period. This was grounded on previous research in cognitive psychology which suggests that more exposure to pictures and a brief cool off period helps to recognize pictures better \cite{tversky1975picture, mandler1976some, Pezdek87memoryfor}. 

We hypothesize that cognitive errors in our scheme could be due to three possible reasons: (a) the user confuses some of the decoys as pass-emojis since only a subset of pass-emojis are present in a challenge ($l = 30$), (b) the user makes errors in computing $f$, and/or (c) the number of pass-emojis is high (14). To pinpoint the reason, we asked the user to do the following in order: (a) to identify if confusion arises in recognizing pass-emojis during authentication, we asked the user to authenticate six times into the system simply by \emph{selecting} their pass-emojis present in the challenge (with no weights); (b) to check if performing arithmetic was the issue, we asked the user to authenticate a further six times. This time the emojis had weights and the user had to compute $f$ and then press the button corresponding to the correct response; (c) to find out if the high number of pass-emojis was an issue, we asked the user to selected their 14 pass-emojis from the total pool of 180 and recorded their selection. Notice that this session did not include any biometric study.

\section{Results}
\label{sec:results}

\subsection{Results from Phase 1}
Recall that the goal of Phase 1 of the user study was to decide which symbol set in Table~\ref{tab:symbols} is the best in terms of repeatability and mimicking hardness. We identified two possible scenarios: \textit{best} and \textit{worst} case and select features separately for the two scenarios. The two scenarios differ in the way testing samples are chosen from the user and the attacker. Note that we pair two users who were given the same symbol set to act as an attacker on the other's symbols. Hence we have 10 user-attacker (ordered) pairs for each symbol set. 

\begin{itemize}
\item \emph{best case scenario:} The first 10 biometric samples out of 13 collected in Session 1 from a user are used for training the classifier. The remaining three are used for testing the accuracy of the classifier for the same user. The attacker samples come from his/her last three samples of the training data in Session 1. 
\item \emph{worst case scenario:} The first 10 biometric samples out of 13 collected in Session 1 are used for training the classifier. The three samples from the same user for testing the accuracy of the classifier come from Session 2 (which was a week apart). The three attacker samples come from the samples entered after the video based observation attack in Session 2. 
\end{itemize}

Algorithm~\ref{algo:feat-select} was used to select features for each symbol in the best and worst case scenarios. The selected feature set is the one, which provides the least average FPR. 
\iffulledition
A tie in the average FPR is broken by choosing the feature set corresponding to the smallest value of the threshold parameter $z$. 
\fi
Essentially, features selected in the best case scenario are the best in providing security against random  
\iffulledition
attacks and ensuring a minimal level of consistency between the user's renderings of the same symbol. 
\else
attacks.
\fi
The features selected in the worst case scenario are best in terms of security against video based observation attacks and ensuring consistency between user's renderings even when they are done after a gap in time. 

Table \ref{table:phase1results1} shows the results (note that the TPR is one in all cases). Complex words yield the least FPR in both the best and worst case scenarios. The individual FPR for complex words was:  0.0, 0.06, 0.0, 0.2, and 0.0 for $\mathtt{xman}$, $\mathtt{bmwz}$, $\mathtt{quak}$, $\mathtt{hurt}$ and $\mathtt{fogy}$, respectively, in the worst case scenario. Also, against random attacks, all symbol categories have an almost 0\% FPR. The table also shows the top features in each symbol set. 
\iffulledition
By top we mean those features that are present in the selected feature set of at least two or more symbols in the symbol set. 
\fi
%three symbols in the category, with the exception of the best case scenario for complex words where the top features appeared in a maximum of two words. 
Simple features such as coordinates, change in coordinates and stylometric features such as height, area, angles and margins constitute the features that provide repeatability and hardness of mimicking across all symbol sets. 
%In fact, the top five features in the best case scenario are: $\mathtt{y}, \delta \mathtt{x}, \delta \mathtt{y}, \mathtt{height}, \theta_{\mathtt{slope}}$. Similarly, the top five features in the worst case scenario are: $\delta \mathtt{x}, \delta \mathtt{y}, \mathtt{TMP}, \mathtt{height}, \theta_{\mathtt{path}}$. 
We do not have enough evidence to conclude whether device-interaction features (see Table~\ref{table:features}) are useful in our case.
\begin{table*}[!ht]
\centering
\caption{Results for best and worst case scenarios for different symbol sets.}
\label{table:phase1results1}
\begin{tabular}{r|c|c|c|c}
Symbol set & Average FPR (best case)  &  Average FPR (worst case) & Top features (best case)  & Top features (worst case)\\
\hline\hline
easy words &0.01  & 0.24 & $\mathtt{x}, \mathtt{y}, \delta \mathtt{x}, \delta \mathtt{y}, \mathtt{TMP}, \theta_{\mathtt{slope}}, \theta_{\mathtt{path}}, \mathtt{R}_{\mathtt{x}}$& $ \mathtt{TMP},  \mathtt{height}, \mathtt{WHR},  \theta_{\mathtt{slope}},\theta_{\mathtt{path}}$\\
 complex words & 0.00 &  0.05& $\mathtt{y}, \delta \mathtt{y}, \mathtt{p},\mathtt{height}, \mathtt{area}, \theta_{\mathtt{slope}}, \mathtt{R}_{\mathtt{y}}$& $ \delta \mathtt{x}, \mathtt{height}, \theta_{\mathtt{path}} $\\
  easy figures &0.01& 0.38 & $\mathtt{y}, \delta \mathtt{x}, \delta \mathtt{y}, \mathtt{p}, \mathtt{F}, \mathtt{height}, \mathtt{area}, \theta_{\mathtt{slope}}, \theta_{\mathtt{path}}$& $\mathtt{y},  \delta \mathtt{y}, \mathtt{p}, \mathtt{height} $\\
  complex figures & 0.01 & 0.39& $\delta \mathtt{x}$& $\mathtt{x}, \mathtt{TMP}, \mathtt{BMP}$
\end{tabular}
\end{table*}

To dig deeper into why some symbol sets have poorer average FPR than others in the worst case scenario, we did the following for each symbol: First, we fix $z = 1$, pick best features for each symbol through the feature selection algorithm and pick first 10 biometric training samples from Session 1 to train the classifier for each user. Next, we tested the classifier, (a) using user's own last three samples from Session 1 to obtain TPR values, (b) using user's three samples from Session 2 to obtain TPR values, and (c) using attacker's three samples from Session 2 to obtain FPR values. The results are shown in Table \ref{table:phase1results3}. The average TPR for all users for Session 1 is denoted $\text{TPR}_1$, whereas for Session 2 is denoted $\text{TPR}_2$. We can see that the average TPR for Session 2 decreases from Session 1 for complex figures drastically which means that users find it hard to repeat drawings of complex figures. A near consistent average TPR  but a high average FPR between the two sessions for easy words and easy figures means they are repeatable but not secure against video based observation attacks. The reason for easy words to be easily mimicked is because of the presence of letters, which do not contain many sharp turns such as \emph{o}, \emph{c} and \emph{s}. The easy figures were easily attacked because drawing them does come naturally to the users and hence they draw them slowly, which makes it easy for an attacker to pick and then mimic. The results for complex words show that they are both highly repeatable and cannot be easily mimicked. Users can write words fluently (due to years of practice), thereby making them difficult to be mimicked.  

\begin{table}[!ht]
\centering
\caption{Results indicating repeatability and resilience against observation attacks for different symbol sets.}
\label{table:phase1results3}
\begin{tabular}{r|c|c|c}
Symbol Category & $\text{TPR}_1$ (average)&  $\text{TPR}_2$ (average) & Average FPR \\
\hline\hline
easy words &0.93  & 1.00 & 0.24\\
 complex words & 0.91 &  1.00 & 0.05\\
  easy figures &0.68  & 0.60 & 0.21\\
  complex figures & 0.70 & 0.53& 0.39
\end{tabular}
\end{table}

\subsection{Results from Phase 2}
Recall that the goal of Phase 2 of the user study was to test our complete authentication scheme. We discuss these results in the following.

\subsubsection{Registration Time}
Table~\ref{table:phase2trainresults} shows the time taken by different user groups in completing the training. The average time to select 14 pass-images is around 2 minutes for all groups. The maximum training time is around 12 minutes for Group 3, since it had the most amount of training. 
\iffulledition
Users in both Group 2 and Group 3 spend 50\% of their total training time for the cognitive scheme to familiarize themselves with their pass-emojis and also learning how to use the scheme. The time to collect biometric samples takes 50\%, 30\%, and 37\% of the total training time for Group 1, 2 and 3, respectively. 
\fi
Since registration is one-off we do not consider the amount of time taken as a major hurdle, specially when most of the users reported enjoying the registration phase as it had a game-like feel to it (see Section~\ref{subsub:question}). 

%As, expected the total training time for Group 3 is maximum at 12 minutes and 26 seconds. 

\begin{table}[!ht]
\centering
\caption{Average registration time (in seconds) of different user groups in Phase 2.}
\label{table:phase2trainresults}
\begin{tabular}{c|c|c|c|c}
 \multirow{ 2}{*}{Group} & Pass-emojis  &   Cognitive  &  Biometric &  Total  \\
 & selection time & training time & training time  & training time \\
\hline\hline
  1&128 & 0 & 129&257\\
  2  & 114 &  284 &174& 573\\
  3  &105  & 359 & 282 &746
\end{tabular}
\end{table}

\subsubsection{Authentication Time}
Table \ref{table:phase2loginresults} shows the average authentication time (per round) taken by different user groups in the two sessions. Since we were interested in per round statistics, we allow the server to accept or reject after each round. 
\iffulledition
Cognitive time represents the time taken by the user to recognize their pass-images and compute $f$. The time spent on writing the complex words on the touchscreen corresponding to the response is the biometric time. The time between the user's submission of the biometric response and the server's accept/reject message is the processing time. 
\fi
Generally, the user spends 15-20 seconds in computing $f$ and 6-8 seconds in entering the biometric response. The results show that more training helps the users to recognize their pass-emojis quicker since Group 3 has the least login time among all the three user groups in Session 1. We also see that the average total time for authentication does not change drastically between the two sessions. 

\begin{table*}[!ht]
\centering
\caption{Authentication statistics for different user groups.}
\label{table:phase2loginresults}
\begin{tabular}{c|c|c|c|c|c|c|c}
\multirow{2}{*}{Group \& session} & Av. cognitive &  Av. biometric & Av. processing & Av. total & Success & Cognitive &  Biometric \\
& time (sec)  & time (sec) & time (sec) & time (sec) & rate (\%) & errors (\%) & errors (\%)\\
\hline\hline
 Group 1 - Session 1 (Phase 2)&18.3 &  7.9 &0.7& 27.0& 38.3&31.6&31.0\\
 Group 2 - Session 1 (Phase 2) & 19.8 &  6.4 &0.7& 27.0& 50.0&18.3&36.0\\
 Group 3 - Session 1  (Phase 2)&12.2  & 5.6 & 0.8 &18.7&85.0&15.0&0.0\\ 
  Group 1 - Session 2 (Phase 2)&18.5 &  7.5 &0.7& 26.8&26.6&55.0&18.3\\
 Group 2 - Session 2  (Phase 2)& 18.4 &  6.4 &0.7& 25.6& 23.3&55.0&26.6\\
 Group 3 - Session 2 (Phase 2)&15.8  & 5.4 & 0.9 &22.0&50.0&41.6&8.3\\
 Group 3 - Session 1 (Phase 3)& -  & - & - &-&94.0&6.0&-\\
 Group 3 - Session 2 (Phase 3)& -  & - & - &-&86.0&14.0&-

\end{tabular}
\end{table*}

\subsubsection{Authentication Errors} 
Table \ref{table:phase2loginresults} also shows the percentage of successful authentication attempts along with the cognitive and biometric errors. 
%which contributed to failed attempts for different user groups across two sessions. 
Recall that each user attempted six authentications in each session. So, we have a total of $v = 60$ authentication attempts for each user group in each session. If the users were randomly submitting a cognitive response, the probability that $i$ out of the $v$ cognitive attempts would succeed is given by 
\iffulledition
\[
p \doteq \binom{v}{i} p_{\text{RG}}^i  (1 - p_{\text{RG}})^{v - i}.
\]
\else
$p \doteq \binom{v}{i} p_{\text{RG}}^i  (1 - p_{\text{RG}})^{v - i}$.
\fi
We will consider $i \ge 20$ out of the $60$ attempts ($< 66 \%$ error rate) as statistically significant because then we have $p < 0.05$. This would imply that the users were not successfully passing a cognitive challenge just due to luck. By looking at the table, we see that all groups had cognitive error rates far less than this. We also see that cognitive training aids the user's short term memory. Users from Group 3 could authenticate successfully 85\% of the time (9 out of the 10 users could authenticate 5 or more times). Users in Group 1 who did not have any cognitive training successfully authenticated only 36\% of the time. Group 2 users, who received a shorter cognitive training than Group 3, accrue 18\% cognitive errors, similar to the cognitive errors from Group 3. However, as the number of biometric training samples collected for users in Group 2 is 5 as opposed to 10 from Group 3, most failures originate from biometric errors. Hence, we believe that if the same number of biometric training samples were collected from users in Group 2 as in Group 3, the performance would be similar to the latter group. In other words, cognitive training time could be 
\iffulledition
reduced. We see
\else
reduced. We see
\fi
a drastic decrease in the successful authentication attempts in Session 2 from Session 1 especially for Group 3 (from 85\% to 50\%) and Group 2 (from 50\% to 24\%). Cognitive errors are predominantly responsible for the drastic decrease as they caused more than half of the authentication attempts to fail for Group 2 and 3, and 40\% for Group 1. To find out the actual cause for such a high number of cognitive errors, we did Phase 3 of the study, whose results are described in Section~\ref{sub:results:phase3}.

\subsubsection{Attack Statistics}
We picked those 12 users to be attacked who successfully authenticated 5 out of 6 times in Session 1 from all the 30 users. Nine users belong to Group 3; two users belong to Group 2 and one user belong to Group 1. Each user among the 30 users in the three groups attacks only one of the 12 target users and performs three random and three video based observation attacks. Hence the total number of random and video based observation attacks is 90 each. If we denote the average FPR of our system by $\overline{\text{FPR}}$ then the probability of a random attack can be estimated as
\iffulledition
\[
p_{\text{tot}} = p_{\text{RG}} \times \overline{\text{FPR}} \approx 0.256 \times 0.05 \approx 0.013.
\] 
Thus $i$ out of $v = 90$ correct guesses would be binomially distributed as
\[
p \doteq \binom{v}{i} p_{\text{tot}}^i (1 - p_{\text{tot}})^{v - i}.
\]
\else
$
p_{\text{tot}} = p_{\text{RG}} \times \overline{\text{FPR}} \approx 0.256 \times 0.05 \approx 0.013
$. Thus $i$ out of $v = 90$ correct guesses would be binomially distributed as
$
p \doteq \binom{v}{i} p_{\text{tot}}^i (1 - p_{\text{tot}})^{v - i}
$.
\fi
Thus we will consider an $i \ge 4$ as statistically significant since then $p < 0.05$. The percentage of successful attacks in our study was 3.33\% (3 times) for both random and video based observation attacks (which is not statistically significant), and none of the successful attacks were consecutive. In all six cases, we found that the target user writes the words using block letters and not cursively. This makes it easier for an attacker to mimic the user~\cite{capital}. 
%All the attacks against the target user who wrote cursively are not successful in our user study. 

\subsubsection{Distribution of Symbols in the Empty Case}
In the empty case, the user is supposed to write a random complex word. We are interested in finding if the resultant distribution of symbols is random or not. We had a total of $v = 34$ instances of the empty case. The probability of randomly choosing a word is $\frac{1}{d} = \frac{1}{5}$. The number of times, $i$, a word was written in a total of $v$ empty cases, is once again binomially distributed (conditioned on the null hypothesis) with this probability. We will consider $i \le 3$ and $i \ge 11$ as statistically significant (as they imply $p < 0.05$). We found that the word $\mathtt{xman}$ (corresponding to $r = 0$) was significantly overused ($i = 13$), whereas the word $\mathtt{bmwz}$ (mapped to $r = 1$) was significantly less used ($i = 2$). The frequency of occurrence of other words did not deviate (statistically) significantly. We believe the reason for overuse of $\mathtt{xman}$ might be because the user thought that an empty case implies the cognitive response is 0. Note that the users were told that they need to write \emph{any} word in the empty case. 

%\begin{table}[!ht]
%\centering
%\caption{Distribution for Symbols in Zero Case Scenario}
%\label{table:zerocaseresults}
%\begin{tabular}{l|p{1cm}|p{1cm}|p{1cm}|p{1cm}|p{1cm}}
%Number of Times & 0 & 1 &2 &3 &4 \\
%\hline\hline
%34&38\%  &  6\%&14\%&17\%&23\%
%\end{tabular}
%\end{table}

\subsubsection{Effect of Number of Pass-Emojis Present}
From Table~\ref{table:phase2secretsresults} we see that the percentage of cognitive errors and authentication time steadily increases with an increasing number of pass-emojis present in the challenge, which is what we would expect. 
%On the other hand, the number of pass-images present also has an effect on the authentication time, increasing from about 14.63 seconds when the number of pass-images is one to 22.60 seconds when five of the pass-images are present. 
Surprisingly, the time taken in the empty case is more than the time taken when one or more pass-emojis is present. A likely reason for this is that the user needs more time to ensure if it is indeed the empty case. 

 \begin{table}[!ht]
\centering
\caption{Time taken and percentage of cognitive errors a function of number of pass-emojis present in a challenge.}\label{table:phase2secretsresults}
\begin{tabular}{c|c|c|c}
\# of pass-emojis & Frequency &  Average time (sec) & Cognitive errors (\%)\\
\hline\hline
0&34&18.08&0.00\\
1&83&14.63&36.14\\
2&111&17.52&42.34\\
3&67&16.90&44.77\\
4&40&18.30&50.00\\
5&21&22.60&52.38\\
6&4&16.22&25.00
\end{tabular}	
\end{table}
%\item Time spend in calculating cognitive response as  function of combination of weights (sum) in secret images.
%\item Errors done in calculating cognitive response as  function of combination of weights (sum) in secret images.

\subsubsection{Pass-Emojis Chosen by Users}
The probability that an emoji is present in a random sample of $k$ emojis out of $n$ is $\frac{k}{n}$. Thus in $v$ random samples, the probability that an emoji occurs $i$ times is given by 
\iffulledition
\[
p \doteq \binom{v}{i} \left( \frac{k}{n} \right)^i \left( 1 - \frac{k}{n} \right)^{v - i}.
\] 
\else
$p \doteq \binom{v}{i} \left( \frac{k}{n} \right)^i \left( 1 - \frac{k}{n} \right)^{v - i}$.
\fi
We had 30 users in Phase 2. Thus, setting $v = 30$ in the above, we see that if an emoji occurs $i \ge 6$ times in the 30 chosen pass-emojis, we will consider the event statistically significant ($p < 0.05$).\footnote{For the lower tail, we see that the probability is always higher than $0.05$ since $v$ is small.}  Our results show that 15 emojis were selected by at least six or more users. These are shown in Table~\ref{table:popemojis} along with the number of users who chose them. Ten of the 15 emojis are animals, which seems to indicate that users were choosing their pass-emojis using an animal theme. This is perhaps also due to the fact that animals constituted a high percentage of the total emojis. 
%\textcolor{red}{TODO: find the icons that occur $i \ge 6$ times. Show the top 10 icons (if they all occur $\ge 6$ times). This will replace Figure~\ref{figure:popularimages}.}

%Our results show that users make themes to pick their pass-images. The top 15 images chosen by all the  users as their pass-images is shown in Table~\ref{table:popemojis}. As top 10 pass-images include 8 animals, it is expected that animal would be the highest chosen category among the users. We find that the 5 top categories constitute 53\% of the total chosen images as shown in Table \ref{table:secretimagesresults}. The other popular categories are food, fruits, transport and sports. 

%\begin{table}[!ht]
%\centering
%\caption{Top 5 pass-images  Category Distribution}\label{table:secretimagesresults}
%\begin{tabular}{c|c}
%Category & Percent of Total Images\\
%\hline\hline
%Animal&25.4 \\
%Food&8.3\\
%Fruit&8.1\\
%Transport&5.9\\
%Sports&5.2
%\end{tabular}
%\end{table}
\iffulledition
\renewcommand{\mysize}{0.45cm}
\else
\renewcommand{\mysize}{0.35cm}
\fi
\begin{table}[!ht]
	\begin{center}
    \begin{tabular}{ c | l }
		Frequency & Emojis \\ \hline\hline
		9 & \raisebox{-.5\height}{\includegraphics[width=\mysize]{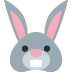}}  \\
		8 & \raisebox{-.5\height}{\includegraphics[width=\mysize]{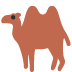}} \raisebox{-.5\height}{\includegraphics[width=\mysize]{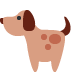}} \raisebox{-.5\height}{\includegraphics[width=\mysize]{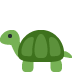}} \\
		7 & \raisebox{-.5\height}{\includegraphics[width=\mysize]{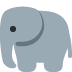}  \includegraphics[width=\mysize]{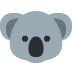}  \includegraphics[width=\mysize]{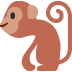}} \\
		6 & \raisebox{-.5\height}{\includegraphics[width=\mysize]{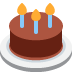} \includegraphics[width=\mysize]{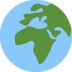} \includegraphics[width=\mysize]{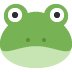} 
		\includegraphics[width=\mysize]{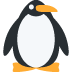} \includegraphics[width=\mysize]{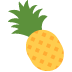} \includegraphics[width=\mysize]{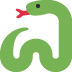} 
		\includegraphics[width=\mysize]{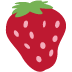} \includegraphics[width=\mysize]{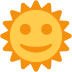}}
    \end{tabular}
    \caption{The 15 most popular emojis in users' pass-emojis.}
    \label{table:popemojis}
    \end{center}
\end{table}

%\begin{figure*}[!ht]
%\centering
%
%
%\subfloat[Bunny]{%
%  \includegraphics[width=1.15cm]{figures/bunny}%
%  \label{fig:bunny}%
%}\qquad
%\subfloat[Camel]{%
%  \includegraphics[width=1.15cm]{figures/camel}%
%  \label{fig:camel}%
%}\qquad
%\subfloat[Dog]{%
%  \includegraphics[width=1.15cm]{figures/dog}%
%  \label{fig:camel}%
%}\qquad
%\subfloat[Turtle]{%
%  \includegraphics[width=1.15cm]{figures/turtle}%
%  \label{fig:camel}%
%}\qquad
%\subfloat[Elephant]{%
%  \includegraphics[width=1.15cm]{figures/elephant}%
%  \label{fig:camel}%
%}\qquad
%\subfloat[Koala]{%
%  \includegraphics[width=1.15cm]{figures/koala}%
%  \label{fig:bunny}%
%}\qquad
%\subfloat[Monkey]{%
%  \includegraphics[width=1.15cm]{figures/monkey}%
%  \label{fig:camel}%
%}\qquad
%\subfloat[Cake]{%
%  \includegraphics[width=1.15cm]{figures/cake}%
%  \label{fig:camel}%
%}\qquad
%\subfloat[Earth]{%
%  \includegraphics[width=1.15cm]{figures/earth}%
%  \label{fig:camel}%
%}\qquad
%\subfloat[Frog]{%
%  \includegraphics[width=1.15cm]{figures/frog}%
%  \label{fig:camel}%
%}
%
%
%\caption{Top 10 most selected images for pass-images}
%\label{figure:popularimages}
%\end{figure*}

\subsubsection{Recognizing Pass-Emojis}
The minimum, maximum and average number of pass-emojis recognized was, respectively, (7, 12, 9.0) for Group 1, (8, 13, 10.5) for Group 2 and (10, 14, 12.1) for Group 3. These results were obtained by asking the users to select their pass-emojis from the total pool of emojis after a gap of one week.
%Table \ref{table:recallresults} shows the number of pass-emojis users from different categories could recognize after a period of one week. These results were obtained by asking the users to select their pass-emojis from the total pool of emojis. 
If the user does not remember any of the pass-emojis, the probability of correctly selecting $i$ out of $k$ emojis is given by
\iffulledition
\[
p \doteq \frac{\binom{k}{i}\binom{n - k}{k - i}}{\binom{n}{k}}.
\] 
\else
$
p \doteq {\binom{k}{i}\binom{n - k}{k - i}}/{\binom{n}{k}}
$.
\fi
An $i \ge 4$ is significant (since $p < 0.05$). We can see that all groups were able to remember a significant number of their pass-emojis. 
%Users from Group 3 (who get the most training) can recognize on average 12 of their pass-emojis, while users from Group 1 could only recognize 9 on average. 
The results indicate that more training may help users in recognizing their pass-emojis in the longer term. However, the higher recognition rates for both Groups 2 and 3 do not translate into higher successful authentication attempts in Session 2. We look into more details of why this happens in the results for Phase 3. 
%The minimum number of pass-emojis a user could recognize is seven. 
%The literature \cite{miller1956magical, cowan2010magical} says that most adults can store between 5 to 9 items in their working memory. 
We can also conclude from this that without much training, users may easily recognize around up to 7 emojis even after a gap in time. Recognizing 10 or more emojis in the longer term requires more training.

%\begin{table}[!ht]
%\centering
%\caption{Pass-emojis recognition stats.}
%\label{table:recallresults}
%\begin{tabular}{c|c|c|c}
%Group & Minimum \#&  Maximum \# & Average\\
%\hline\hline
% 1 &7 & 12 & 9\\
% 2  & 8 &  13  &10.5\\
% 3  &10  & 14 & 12.1
%\end{tabular}
%\end{table}

\subsubsection{Guessing Pass-Emojis}
From the analysis above, we see that if an attacker can guess more than $4$ pass-emojis of the target user, we will consider that the attacker has significant advantage over random guess. We found that five of the 30 attackers were able to guess 4 or more pass-emojis of the target user, and one attacker guessed as many as 11. In the last case, the attacker thought that the target user might have picked pass-emojis according to a theme comprising of animals. This result shows that picking pass-emojis based on a theme might lead to more chances of being attacked.

%\begin{table}[!ht]
%\centering
%\caption{Secret Image Guessed by an Attacker}
%\label{table:guessresults}
%\begin{tabular}{p{1.5cm}|p{1.5cm}|p{1.5cm}}
%Minimum \#&  Maximum \# & Average\\
%\hline\hline
% 0& 11 & 1.73
%\end{tabular}
%\end{table}

\subsubsection{ Questionnaire Statistics} 
\label{subsub:question}
At the end of Session 2, we asked the (30) users to fill a questionnaire on a Likert scale of 1 to 5, where 1 means Strongly disagree, 2 means Disagree, 3 means Neither Agree or Disagree, 4 means Agree and 5 means Strongly Agree. The general consensus about the ease of writing words on the smartphone screen was a rating of 4. 
%and if the user can repeatedly write the words consistently easily are 4 (Agree). 
The users liked playing the training game during the registration with an overall rating of 4. The overall usability of the scheme received mixed rating from the users (3). However, the users who mostly rated 1 or 2 for usability said that they are likely to use the system if it can provide a high security guarantees. The major issue the users had with our scheme is the number of pass-emojis. The rating was 2 when the users were asked if they could manage 14 pass-emojis easily. A 53\% of the users say that they would not like to use the system because of 14 pass-emojis, and 30\% users found it hard to recognize their pass-emojis during authentication.
% These percentages indicate as to why we  get a very high number  of failed login attempts in Session 2.  
Only 16\% and 6\% of the users complained about entering biometric responses and computing $f$.  

Most users (53\%) prefer 6-10 pass-emojis as their secret followed by 30\% who prefer no more than 5 pass-emojis. 
%The rest 17\% of the users respond that they prefer more than 11 pass-images as their password. 
In response to the question on the size of $l$ (i.e., the window size), 53\% of the users responded with 0-10 emojis.  23\% users said 11-20 and a similar percentage were fine with the current scheme (21-30 emojis). When asked how they picked their pass-emojis, 18 users said they created a certain theme to make it easy for them to remember their pass-emojis. Some users used multiple themes; 7 users said that they picked animals, 6 users picked food items, 2 users picked tools, 2 users picked sports, one picked recreation and one picked faces. Two users created a theme based on a story. One story was: ``Santa watching sports while eating a lot of food.''
\subsection{Results from Phase 3}
\label{sub:results:phase3}
Recall that this phase was carried out to find the main cause of cognitive errors and to improve our training to alleviate the issue. The users did 12 authentication attempts each in Sessions 1 and 2. The first 6 involved merely selecting the pass-emojis present whereas the second involved computing $f$ as well. The results are shown in the last two rows of Table \ref{table:phase2loginresults}. The results show that our improved training module (more exposure to each individual pass-emojis followed by blank screens) drastically decreases the error rate. Even after a week's gap the success rate is 86\%.
\iffulledition
We rule out the possibility that high number of pass-emojis is the most probable reason for high cognitive errors because the minimum, maximum and average number of pass-emojis users can recognized was, (12, 14, 13.6) for Session 1 and (11, 14, 13.5) for Session 2 respectively. In fact, seven out of 10 users can recognize all of their pass-emojis in both the sessions. 
\else
We rule out the possibility that the errors in Phase 2 were due to the size of the secret, as the average number of pass-emojis recognized by the users in Sessions 1 and 2 were 13.6 and 13.5, respectively. 
\fi
We also counted the total number of errors made by the users in the first 6 authentication attempts, which turned up 13, and the last 6 authentication attempts, which turned up 11, adding results from both sessions. This shows no evidence that computing $f$ was causing errors. We, therefore, believe that the main cause of errors is due to the user confusing decoy emojis as his pass-emojis since only a subset of the $k$ emojis are present in the challenge (due to $l$).

\section{Related Work}
\label{sec:rw}
Although our idea is generic enough to use any cognitive authentication scheme, we proposed a new scheme since existing schemes did not possess all the attributes we desired. The low complexity cognitive authentication scheme (CAS) proposed by Weinshall~\cite{CAS} could have been a natural candidate. Unfortunately, it is susceptible to SAT solver based attacks shown by Wagner and Golle~\cite{wagner}. 
\iffulledition
Although they reported that their attack could find the secret after observing 60 rounds, it may be possible to further reduce this with a trade off with time complexity.  
\fi
Furthermore, the CAS scheme 
\iffulledition
uses parameter sizes of $n = 80$ and $k = 30$, and all $n$ images need to be shown on the screen at once, which is hard to display on touch screens of smartphones. 
\else
requires all $n = 80$ images to be shown at once, which is hard to do on small screens.
\fi
The Asghar, Pieprzyk and Wang (APW) scheme~\cite{APW} suffers from the same issue where all $n$ images need to be displayed. The scheme from Li and Teng~\cite{Li1999} requires the user to remember three different secrets and perform lexical-first matching on the challenge to obtain hidden sub-sequences. The cognitive load of their scheme seems high, and it is unclear if a graphical implementation is possible. 
%Although they reported that their attack could find the secret after observing 60 rounds, it may be possible to further reduce this with a trade off with time complexity. 
The HB protocol from~\cite{Hopper} is another candidate which can be modified to use window based challenges, but it requires the user to add random responses with a probability $\eta < \frac{1}{2}$. It is unclear how the user could mentally generate a skewed probability $\eta$. The Foxtail protocol from Li and Shum~\cite{sechci} reduces the response space to $\{0, 1\}$ which means that more number of rounds would be needed. Schemes such as predicate-based authentication service (PAS)~\cite{PAS} only resist a very small number of authentication sessions ($< 10$)~\cite{pas-security}. The convex hull click (CHC) scheme 
\iffulledition
uses a somewhat different approach, in which the response could be any point on the screen. The user locates at least three of his/her pass-images in the challenge and clicks randomly within the imaginary convex hull of the pass-images. 
\else
asks the user to locate at least three pass-images in the challenge and click randomly within the imaginary convex hull of the pass-images. 
\fi
With the default parameter sizes $k = 5$ and $l = 82$ (on average) the scheme is vulnerable to statistical attacks~\cite{yan2012limitations, chc-attack}. Setting aside the issue of how the response space from CHC can be matched to behavioural biometrics, it appears that increasing parameter sizes (to make the scheme more secure) will require higher number of rounds to resist random guesses. Blum and Vempala~\cite{blum2015publishable} propose several simple cognitive schemes which are easy to compute for humans and require little training. Although their schemes are information theoretically secure, the guarantee is only for a small number of observed sessions (6 to 10). The scheme from Blocki et al~\cite{blocki} is provably secure against statistical adversaries, and can resist a sizeable number of observed sessions. The scheme's main drawback is the extensive training which requires a human user to memorize random mappings from 30-100 images to digits, which, even with memory aids such as mnemonics, could take considerable time. An interesting open question is to see if their proof strategy can be extended to show if BehavioCog is secure against statistical adversaries.

A number of touch-based behavioural biometric schemes have been proposed for user authentication ~\cite{xu-soups,touchalytics,unobserve-ndss,shahzad-svde}. Most of these schemes rely on simple gestures on smartphones such as swipes. 
We have argued that if we were to use simple gestures then a much larger number of them need to be accumulated to get good accuracy. Also, swipes are prone to observation attacks~\cite{hassan-khan}.
%A recent work \cite{hassan-khan} also suggest that swipes are highly prone to observation attacks, which doe not make them suitable for our scheme. 
The work of Sherman et al.~\cite{Sherman} does indeed include more complex (free-form) gestures and partly inspired our symbol set of complex figures. However, their gestures are only known to resist shoulder-surfing 
\iffulledition
attacks and not video based observation attacks where the attacker has full control over the playback.
\else
attacks.
\fi
The closest work similar to ours is by Toan et. al. \cite{Nguyen}. Their scheme authenticates users on the basis of how they write their PINs on the smartphone touch screen using $x, y$ coordinates. In comparison, we do a  more detailed feature selection process to identify features, which are repeatable and resilient against observation attacks. Furthermore, they report an equal error rate (EER) of 6.7\% and 9.9\% against random and shoulder-surfing attacks, respectively. Since these are EER values, the TPR is much lower than 1.0. To obtain a TPR close to 1.0, the FPR will need to be considerably increased. Thus, after observing one session, the observer has a non-negligible chance of getting in (since the PIN is no longer a secret). To achieve a low probability of random guess, the number of rounds in their scheme would need to be higher. Furthermore, after obtaining the PIN, the attacker may adaptively learn target user's writing by querying the authentication service. The use of a cognitive scheme, as mentioned before, removes this drawback. Another work close to ours is KinWrite~\cite{kinwrite}, which asks the user to write their passwords in 3D space, and then authenticates them based on their writing patterns in 3D space. We note that KinWrite also suffers from the same drawbacks mentioned before. Pure graphical password schemes such as D{\'e}J\`{a} Vu~\cite{Dhamija} and Draw-a-Secret scheme~\cite{DAS} where the user has to click directly on pass-images or reproduce the same drawing on the screen, have the same vulnerability. A single observation reveals the secret, and if a behavioural biometric component is added, it is as good as using it as a standalone system.  %Furthermore, they report an equal error rate (EER) of 6.7\% and 9.9\% against random and shoulder surfing attacks, respectively. Since these are EER values, the TPR is much lower than 1.0. To obtain a TPR close to 1.0, the FPR will need to be considerably increased. Thus, after observing one session, the observer has a non-negligible chance of getting in (since the PIN is no longer a secret). To achieve a low probability of random guess, the number of rounds in their scheme would need to be higher. Furthermore, after obtaining the PIN, the attacker may adaptively learn target user's writing by querying the service. The use of a cognitive scheme, as mentioned before, removes this drawback. KinWrite~\cite{kinwrite} is another work closely related to ours which asks the user to write their passwords in 3D space, and then authenticates them based on their writing pattern in 3D space. We note that KinWrite also suffers from the same drawbacks mentioned before. Pure graphical password schemes such as D{\'e}J\`{a} Vu~\cite{Dhamija} and Draw-a-Secret (DAS) scheme~\cite{DAS} where the user has to click directly on pass-images or reproduce the same drawing on the screen, have the same vulnerability. A single observation reveals the secret, and if a behavioural biometric component is added, it is as good as using it as a standalone system. 

\section{Discussion and Limitations}
\label{sec:discuss}
\iffulledition
To begin, our initial user study showed that a large number of pass-emojis is hard to recognize, especially when the protocol uses a window based challenge (a random subset of total objects). This was most likely due to users confusing emojis similar to their pass-emojis as their own (e.g., having similar color) since not all pass-emojis are present to clear the confusion and/or the exact number of pass-emojis present is unknown. Unfortunately, fixing the number of pass-emojis in a challenge is not an answer as this makes the scheme susceptible to statistical attacks as discussed in Section~\ref{sub:fa}. 
We 
\else
To begin, we
\fi
have shown that a carefully designed training module based on results from cognitive psychology helped users recognize their pass-emojis better. The potential of this needs to be further explored to see how large a set of images could be successfully recognized by users after longer gaps. We reiterate that our scheme also allows a smaller number of pass-emojis but at the  expense of withstanding a smaller number of observed authentication sessions. This is not a serious drawback for authentication on smartphones, as it may be impractical for an attacker to follow a mobile user over a sustained period to record enough observations.

Turning to behavioural biometrics, we observe a slight increase in biometric errors (Group 3 in Table~\ref{table:phase2loginresults}) after a one week period. A remedy is to frequently update the biometric template by replacing older samples with samples from successful authentication attempts~\cite{chauhan2016gesture}. In general, behavioural biometrics tends to evolve over time. On the flip side this very disadvantage is desirable from a privacy perspective. Once stolen, physiological biometrics such as fingerprints cannot be replaced. If the template of a word is stolen, the situation is less dire; the user behaviour might evolve over time or the word itself could be replaced. This is one reason for our preference of behavioural biometrics. Changing topic, the skeptic may object to our claim that our scheme resists observations, because while the cognitive scheme can in theory be proven to resist observation attacks, a similar claim for the behavioural biometric scheme has only been empirically demonstrated through our user study. We somewhat agree; although the evidence provided by us points to the plausibility of our claim, the exact difficulty in mimicking cursively written words derived from certain English letters (see Section~\ref{sub:symbols}) needs to be further explored.

% and, if at all possible, should be proven to hold. , trading with an increased average authentication time

Looking at Table~\ref{table:phase2loginresults}, our cognitive scheme might be susceptible to timing attacks~\cite{timing}, in which the attacker can guess how many pass-emojis are present based on the time the user takes to reply to a challenge. One way to circumvent this is to not allow the user to proceed unless a fixed amount of time has elapsed based on the highest average-time taken per number of pass-emojis present. We also did not consider an adversary that intercepts biometric samples en route from the smartphone to the authentication service $\mathcal{S}$. An obvious solution is to use an encrypted channel. A final remark is that in order to protect the user's secret (pass-emojis and biometric templates), the authentication service could keep it encrypted and decrypt it only briefly during authentication. To ensure that the protocol can be carried out without ever decrypting the secret requires further research into application of techniques such as fuzzy vaults~\cite{fuzzy-vault} and functional encryption~\cite{fe}. For now, we leave it as future work. 

\section{Conclusion}
\label{sec:conclude}
The promise offered by cognitive authentication schemes that they are resistant to observation has failed to crystallize in the form of a workable protocol. Indeed, many researchers have come to the conclusion that such schemes may never be practical. We do not refute this, but instead argue that combining cognitive schemes with other authentication schemes may make the hybrid scheme practical and still resistant to observation, albeit in a more empirical sense. Our approach was to seek help from touch-based behavioural biometric authentication. We demonstrate considerable gain in usability over standalone cognitive schemes, but concede that several aspects need to be improved in future.  Our proposed cognitive and behavioural biometric schemes are not the only ones possible. In fact, we need not confine ourselves to touch based biometrics, and may explore other behavioural biometric modalities. This way, several different constructions are conceivable.  




%\bibliographystyle{IEEEtranS}
\small

\bibliographystyle{IEEEtranS}
{
\bibliography{IEEEabrv,bibliography}}
%\end{b}

\end{document}